\documentclass[onecolumn,aps,nofootinbib,preprint]{revtex4}
\pdfoutput=1

\usepackage{graphicx}
\usepackage{amsmath}
\usepackage{hyperref}
\usepackage{amssymb}
\usepackage{color}
\usepackage{leftidx}
\usepackage{multirow}
\usepackage{siunitx}

\newcommand{\nn}{\nonumber}
\newcommand{\beq}{\begin{equation}}
\newcommand{\eeq}{\end{equation}}
\newcommand{\bea}{\begin{eqnarray}}
\newcommand{\eea}{\end{eqnarray}}

\def\Mpc{\ensuremath\mathrm{Mpc}}

\def\s8{\ensuremath{\sigma_8}}
\def\h0{\ensuremath{H_0}}
\def\dN{\ensuremath{\Delta N_{\mathrm{fluid}}}}
\def\g0{\ensuremath{\Gamma_0}}

\def\eq{\ensuremath{\mathrm{eq}}}
\def\dec{\ensuremath{\mathrm{dec}}}

\def\mH{\ensuremath\mathcal{H}}
\def\mG{\ensuremath\mathcal{G}}
\def\a{\ensuremath\alpha}
\def\ok{\ensuremath\kappa}

\def\LC{\ensuremath{\Lambda \mathrm{CDM}}}
\def\LCs{\ensuremath{\Lambda \mathrm{CDM} \ }}

\def\IDS{\ensuremath\mathrm{IDS}}

\def\cdm{\ensuremath\mathrm{cdm}}
\def\idm{\ensuremath\mathrm{idm}}

\def\dr{\ensuremath\mathrm{dr}}
\def\ra{\ensuremath\mathrm{r}}
\def\ma{\ensuremath\mathrm{m}}
\def\so{\ensuremath\mathrm{s}}

\def\ocdm{\ensuremath{\omega_{\mathrm{cdm}}}}
\def\ob{\ensuremath{\omega_{\mathrm{b}}}}
\def\ora{\ensuremath{\omega_{\mathrm{r}}}}
\def\oma{\ensuremath{\omega_{\mathrm{m}}}}

\def\odmtot{\ensuremath{\omega_{\mathrm{dm}}^{\mathrm{tot}}}}
\def\oidm{\ensuremath{\omega_{\mathrm{idm}}}}
\def\csp{\ensuremath{c_{\mathrm{sp}}}}

\def\f0{\ensuremath{f^{(0)}}}

\def\df0{\ensuremath{\frac{\partial \f0_\dr}{\partial q}}}
\def\dlnf0{\ensuremath{\frac{\partial \ln \f0_\dr}{\partial \ln q}}}

\def\beq{\ensuremath\longrightarrow\ }

\newcommand{\Sec}[1]{Sec.~\ref{#1}}

\newcommand{\App}[2]{App.~\hyperref[#1]{#2}}
\newcommand{\Fig}[1]{Fig.~\ref{#1}}

\newcommand{\Eq}[1]{Eq.~(\ref{#1})}
\newcommand{\Eqs}[2]{Eqs.~(\ref{#1}) and (\ref{#2})}
\newcommand{\Eqsto}[2]{Eqs.~(\ref{#1})-(\ref{#2})}

\newcommand{\ie}{\textit{i.e.}\ }

\newcommand{\bl}{\left}
\newcommand{\br}{\right}

\newcommand{\ignore}[1]{}

\makeatletter
\def\simgt{\mathrel{\lower2.5pt\vbox{\lineskip=0.5pt\baselineskip=0pt
           \hbox{$>$}\hbox{$\sim$}}}}
\def\simlt{\mathrel{\lower2.5pt\vbox{\lineskip=0.5pt\baselineskip=0pt
           \hbox{$<$}\hbox{$\sim$}}}}
\makeatother

\begin{document}
\setlength{\unitlength}{1mm}

\title{Interacting Dark Sector and Precision Cosmology}

\author{Manuel A. Buen-Abad}
\author{Martin Schmaltz}\email{buenabad@bu.edu and schmaltz@bu.edu}
\affiliation{Physics Department, Boston University; 590 Commonwealth Avenue, Boston, MA 02215, USA}
\author{Julien Lesgourgues}
\author{Thejs Brinckmann}\email{lesgourg@physik.rwth-aachen.de and brinckmann@physik.rwth-aachen.de}
\affiliation{Institute for Theoretical Particle Physics and Cosmology (TTK) RWTH Aachen University, D-52056 Aachen, Germany\vskip.05in}

\begin{abstract}
\vskip.1in

We consider a recently proposed model in which dark matter interacts with a thermal background of dark radiation. Dark radiation consists of relativistic degrees of freedom which allow larger values of the expansion rate of the universe today to be consistent with CMB data ($H_0$-problem). Scattering between dark matter and radiation suppresses the matter power spectrum at small scales and can explain the apparent discrepancies between $\Lambda$CDM predictions of the matter power spectrum and direct measurements of Large Scale Structure LSS ($\sigma_8$-problem). We go beyond previous work in two ways: 1. we enlarge the parameter space of our previous model and allow for an arbitrary fraction of the dark matter to be interacting and 2. we update the data sets used in our fits, most importantly we include LSS data with full $k$-dependence to explore the sensitivity of current data to the shape of the matter power spectrum. We find that LSS data prefer models with overall suppressed matter clustering due to dark matter - dark radiation interactions over $\Lambda$CDM at 3-4 $\sigma$. However recent weak lensing measurements of the power spectrum are not yet precise enough to clearly distinguish two limits of the model with different predicted shapes for the linear matter power spectrum.  In two Appendices we give a derivation of the coupled dark matter and dark radiation perturbation equations from the Boltzmann equation in order to clarify a confusion in the recent literature, and we derive analytic approximations to the solutions of the perturbation equations in the two physically interesting limits of all dark matter weakly interacting or a small fraction of dark matter strongly interacting. 
\end{abstract}

\maketitle

\section{Introduction}\label{sec:int}

Over the last few decades cosmology has reached a level of precision that has allowed scientists to discriminate among the different theories that attempt to explain the Universe's composition, expansion, thermal history, and structure formation. Among these the \LCs paradigm has proved to be in excellent agreement with cosmological data, and has arisen as the ``Concordance" or ``Standard" model of cosmology.

In \LC, dark matter (DM) is made of cold particles whose dominant interactions among themselves and with the rest of the Universe contents (the Standard Model of Particle Physics and the cosmological constant $\Lambda$) is through gravity. This is called CDM, for cold dark matter.

Despite the indisputable success of \LC, cosmological experiments have not unambiguously singled it out as the only acceptable explanation to the observed data. Alternative models similar to \LCs but with different physical properties (v.g. possessing extra relativistic sectors or DM that self-interacts through other forces besides gravity) are allowed within the experimental uncertainties.

In addition to this, recent direct measurements of the Large Scale Structure (LSS) of the universe and specifically of the quantity \s8 (the amplitude of the density fluctuations in spheres with radius of $8 h^{-1} \Mpc$) performed by weak lensing and cluster surveys (v.g. CFHTLenS \cite{Heymans:2013fya}, Planck SZ clusters \cite{Ade:2013lmv,Ade:2015fva}, KiDS \cite{Kohlinger:2017sxk,Joudaki:2017zdt}, and DES \cite{Abbott:2017wau}) return smaller values than that extrapolated from the Planck telescope's cosmic microwave background (CMB) data under the assumption of $\Lambda$CDM. This \s8 tension fluctuates between the 2$\sigma$ and 4$\sigma$ level depending on the data set. Similarly, there is another tension of $\sim 3 \sigma$ between the value of the Hubble Parameter \h0 measured directly \cite{Riess:2016jrr,Bonvin:2016crt} and the smaller one extrapolated from Planck under the $\Lambda$CDM assumption \cite{Ade:2015xua}. Were these discrepancies to be of physical origin (rather than unaccounted-for systematics), they would be a sign of the need for new physics beyond \LC, and that a new cosmological model ought to take its place. Different studies on these discrepancies have appeared in the literature (\cite{Buen-Abad:2015ova,Lesgourgues:2015wza,Poulin:2016nat,MacCrann:2014wfa,Chacko:2016kgg,Canac:2016smv,Bernal:2016gxb,Chudaykin:2016yfk,Archidiacono:2016kkh,Joudaki:2016kym,Lancaster:2017ksf,Oldengott:2017fhy}). 

The fact that CMB experiments such as Planck measure the early (recombination-era) photon anisotropies, whereas the LSS surveys measure the matter perturbations as observed today, suggests that a possible resolution to the tensions in the LSS measurements could come in the form of a relationship between these two kinds of anisotropies that is different to that in \LCs. Since \h0 is correlated with various other cosmological parameters, a modification of the \LCs paradigm could either ease or worsen the tension between CMB data and direct \h0 measurements, depending on the  ingredients of the new model.

Recently, \cite{Buen-Abad:2015ova,Lesgourgues:2015wza,Schmaltz,Chacko:2016kgg} discussed an alternative model to \LCs which can resolve both tensions. This model invokes a dark sector (DS) composed of mutually interacting dark matter (IDM) and dark radiation (DR)\footnote{The possibility of dark matter interacting with dark radiation was first explored in \cite{Boehm:2000gq,Boehm:2003hm}. For other recent work on the subject see \cite{Cyr-Racine:2013fsa,Cyr-Racine:2015ihg}.}. The interactions in the DS act to suppress the Matter Power Spectrum (MPS) with respect to the \LCs case, while the extra relativistic degrees of freedom in the DR act to increase the best-fit value of \h0 from the CMB. The direction of degeneracy in parameter space between (\h0, \s8) or (\h0, $\Omega_\ma$) (where $\Omega_\ma$ is the fraction of today's energy density in the Universe that is made up of non-relativistic matter) is different than in extensions of \LCs with only new massless or light degrees of freedom, $\Delta N$, allowing to resolve both tensions simultaneously, instead of improving one at the expense of the other.  \cite{Lesgourgues:2015wza} also presented a chi-squared fit of the model parameters to cosmological data (Planck CMB, BAO, LSS, and $H_0$). They reported that the data prefer non-zero dark radiation densities and IDM-DR interactions at $~\sim 3 \sigma$ relative to \LCs.  Most of the improvement came from the suppression of large scale structure ($\sigma_8$) in the matter power spectrum while the tension between CMB and direct determinations of $H_0$ was also reduced. Subsequently \cite{Krall:2017xcw} performed a fit of the IDM-DR model which includes pioneering Lyman-$\alpha$ data from the 2004 SDSS \cite{McDonald:2004eu,McDonald:2004xn} and found lower significances for the suppression of the matter power spectrum. However, one might anticipate that a fit to more recent 2016 BOSS Lyman-$\alpha$ data would reverse this trend because the recent BOSS data favors matter power spectra which are consistent with the LSS data included in \cite{Lesgourgues:2015wza}.

Here we perform a new precision fit of the IDM-DR model to cosmological data where we extend the previous work in two important ways: 
\begin{enumerate}
\item We consider a generalization of the IDM-DR model in which we allow for 2-component dark matter. One component is ordinary non-interacting CDM whereas the other is IDM, i.e. cold dark matter which interacts with the DR. This generalization allows for qualitatively different limits which both suppress the matter power spectrum and solve the $\sigma_8$ problem. One can either have all of the DM interact very weakly with the DR  \cite{Buen-Abad:2015ova} or have very little IDM but with strong couplings to the DR so that they form a tightly coupled ``dark plasma" \cite{Schmaltz,Chacko:2016kgg}. These two different limits of the general interacting dark sector model (from now on, IDS model) predict distinct shapes and cosmological time dependences for the matter power spectrum.  
\item In our previous fits large scale structure was only included in the guise of the parameter $\sigma_8$. Here we include the full shape information of the matter power spectrum as measured with weak lensing by CHFTLens  \cite{Heymans:2013fya} and using Luminous Red Galaxies as tracers for LSS by SDSS-DR7 \cite{Reid:2009xm}. 
\end{enumerate}
Including experimental input on the matter power spectrum shape is especially interesting as it has the potential to differentiate between models which are consistent with the same values of $\sigma_8$ but predict a different time (cosmological redshift $z$) and scale (wave number $k$) dependence of the linear matter power spectrum. Our result is that current LSS data is starting to become sensitive to the shape of the matter power spectrum but that the differences in $\chi^2$ are not yet very significant. Clearly, this is an exciting area to watch for future theoretical and experimental developments as the full $k$ and $z$ dependent matter power spectrum carries a lot of information about the cosmological history of the universe and especially the properties of DM.  

This paper contains some analytical results on the calculation of the matter power spectrum as well as numerical results from our fits to data performed with CLASS \cite{Blas:2011rf} and MontePython \cite{Audren:2012wb}. Readers only interested in only one or the other are encouraged to skip to the relevant Sections.  
In \Sec{sec:ids} we review the IDS model as a cosmological model with its new parameters (the effective number of relativistic degrees of freedom, the fraction of the DM which is interacting, and the IDM-DR coupling strength). We give the differential equations for the linear evolution of cosmological perturbations and find approximations to them in the two limits (weakly coupled and dark plasma). In \Sec{sec:mps-cmb-lens} we analyze the effects that the IDS model has on the MPS and CMB spectra, and compare them to the \LCs case. We demonstrate and support our findings with a number of plots generated with CLASS showing the spectra as functions of model parameters.  \Sec{sec:results} contains our fits to data. We list the CMB, LSS and BAO (baryon acoustic oscillations) experimental data which are included in the fits. We show plots of likelihood contours in model parameter space indicating the preferred regions of the IDS model. We also give best fit values and confidence intervals for each model parameters and demonstrate the degree of improvement in the fit for each independent set of experimental data. \Sec{sec:conc} contains our conclusions, a discussion of recent data for which no likelihoods were available at time of this writing and an outlook to the future with possible extensions of this work.
In \App{appA}{A} we give a detailed derivation of the linear cosmological perturbation equations including general DM-DR interactions from the Boltzmann equations (here we lean heavily on the derivations in the ETHOS paper \cite{Cyr-Racine:2015ihg}). We explain the physical assumptions behind the approximations made and elucidate the origin of a discrepancy in the interaction term of our perturbation equations relative to that found in the published version of \cite{Cyr-Racine:2015ihg}. In \App{appB}{B} we discuss the shape of the MPS on the basis of analytical expressions which we obtained following a method due to Weinberg (\cite{Weinberg:2002kg,Weinberg:2008zzc}) of matching approximate solutions to the perturbation equations.

\section{The Interacting Dark Sector model}\label{sec:ids}
%
\subsection{Ingredients and parameters}\label{subsec:pars}

The generalized interacting dark sector (IDS) model contains the following three ``dark" ingredients (in addition to cosmological constant): {\it i.} a component of ordinary non-interacting CDM, {\it ii.} a second component of interacting dark matter (IDM), and {\it iii.} a component of dark radiation (DR) which the IDM couples to. The dark radiation is assumed to have frequent self-interactions so that does not free-stream but behaves instead  as a perfect fluid. The interactions also ensure the DR fluid maintains local thermal equilibrium (this means that for at each space-time point there exists a reference frame in which the radiation has a thermal distribution function).  
Concrete particle physics models which realize these characteristics can be found in \cite{Buen-Abad:2015ova,Lesgourgues:2015wza,Chacko:2016kgg,Ko:2016uft,Ko:2016fcd,Ko:2017uyb}.

We are interested in suppressing the MPS on length scales corresponding to \s8, but leaving it unchanged on larger scales. Perturbations of size corresponding to \s8 enter the Hubble horizon before matter-radiation equality. Therefore we can accomplish what we want if the interactions between DR and DM are effective throughout Radiation Domination (RD) and shut off after equality. Because the expansion rate of the Universe during RD scales as $H \propto a^{-2}$ we require $\Gamma \propto a^{-2}$ too, where $\Gamma$ is the momentum transfer rate for an IDM particle traveling through a DR medium. This ensures that the interactions remain of equal importance throughout RD and become less relevant during Matter Domination (MD), when they are overcome by the expansion of the Universe in this era, $H \propto a^{-3/2}$. This behavior is realized in the concrete particle physics models discussed in \cite{Buen-Abad:2015ova,Lesgourgues:2015wza,Chacko:2016kgg,Ko:2016uft,Ko:2016fcd,Ko:2017uyb}. The IDS model includes the parameters of \LC, which we denote by $\boldsymbol{\theta}_{\LC} \equiv \{ \ob, \ \ocdm, \ \theta_\so, \ n_\so, \ A_\so, \ \tau_\mathrm{reio} \}$, as well as three more:
\begin{itemize}
	\item $\dN \equiv \frac{\rho_\dr}{\rho_{1\nu}}$: the amount of DR, parameterized as the effective number of extra neutrino families.
	\item $\g0 \equiv \Gamma a^2$: the momentum transfer rate from the IDM to the DR today (at redshift $z=0$).
	\item $f \equiv \frac{\oidm}{\odmtot}$, with $\oidm \equiv \frac{\rho_\idm h^2}{\rho_{\mathrm{crit}}}$, and $\odmtot \equiv \ocdm + \oidm$: the fraction of DM that is IDM (\ie that interacts with the DR).
\end{itemize}
We denote $\{ \dN, \g0, f \}$ by $\boldsymbol{\theta}_{\IDS}$.

\subsection{The linear perturbation equations}
\label{subsec:eqs}

The cosmological linear perturbation equations include those for \LC, with additional equations of motion for the IDM and DR fluid perturbations. The new fluids also contribute to the gravitational potentials in the linearized Einstein equations which we do not show here (but see e.g. \cite{Ma:1995ey}). In the conformal Newtonian gauge, the new fluid equations are
\begin{eqnarray}
	\dot\delta_\idm & = & -\theta_\idm + 3 \dot\phi \label{didm_1} \\
	\dot\theta_\idm & = & -\mH \theta_\idm + k^2 \psi + \mG ( \theta_\dr - \theta_\idm ) \label{thidm_1} \\
	\dot\delta_\dr & = & -\frac{4}{3} \theta_\dr + 4 \dot\phi \label{ddr_1} \\
	\dot\theta_\dr & = & k^2 \bl( \frac{\delta_\dr}{4} + \psi \br) - \mG R (\theta_\dr - \theta_\idm) \ , \label{thdr_1} 
\end{eqnarray}
where the derivatives are with respect to conformal time. We also defined $\mG \equiv a \Gamma = a^{-1} \g0$
and $\mH \equiv a H$, and $R$ is\footnote{$R$ ensures energy-momentum conservation within the IDM-DR system. For a careful derivation of $R$ see \App{appA}{A}.}
\begin{equation}
	R \equiv \frac{3}{4}\frac{\rho_\idm}{\rho_\dr} = \frac{3}{4} \bl(3.046+\dN+\frac{8}{7} \bl(\frac{11}{4}\br)^{4/3} \br) \bl( 1+\frac{\ob}{\odmtot} \br)^{-1} \frac{f}{\dN} \frac{a}{a_\eq} \ . \label{R_def}
\end{equation}
A useful reference value is $\dN =0.4$, $f = 1$, $\odmtot = 0.12$, and $\ob = 0.022$, which gives $R(a_\eq) \approx 12.5$.
We can eliminate $\theta_i$ and obtain the second-order equations:
\begin{eqnarray}
	\ddot{\delta}_\idm + ( \mH + \mG )\dot\delta_\idm & = & -k^2 \psi + 3\ddot{\phi} + 3\mH\dot{\phi} + \frac{3}{4}\mG\dot\delta_\dr \label{didm_2} \\
	\ddot\delta_\dr + \frac{k^2}{3}\delta_\dr + \mG R \dot\delta_\dr & = & \frac{4}{3} \bl( -k^2 \psi + 3 \ddot\phi + \mG R\dot\delta_\idm \br) \ . \label{ddr_2}
\end{eqnarray}
Yet another way to rewrite these equations is by defining $\Delta \equiv \delta_\idm - \frac{3}{4}\delta_\dr$ (note that $\dot\Delta = \theta_\dr - \theta_\idm$):
\begin{eqnarray}
	\ddot\delta_\idm + \frac{R}{1+R}\mH\dot\delta_\idm + k^2 \csp^2 \delta_\idm & = & -k^2 \psi + 3\ddot\phi + \frac{R}{1+R}3 \mH\dot\phi + 3 \csp^2 \bl( \ddot\Delta + \frac{k^2}{3}\Delta \br) \label{didm_3}\\
	3 \csp^2 \bl( \ddot\Delta + \frac{k^2}{3} \Delta \br) + \mG \dot\Delta & = & 3 \csp^2 \bl( 3 \mH \dot\phi - \mH \dot{\delta}_\idm + \frac{k^2}{3} \delta_\idm \br) \label{del_1} \ ,
\end{eqnarray}
\begin{equation}
	\mathrm{with} \quad \csp^2 \equiv \frac{1}{3(1+R)} \ . \label{cs2}
\end{equation}

In the limit of tightly coupled interacting DM and DR, $\mG \gg \mH$, Eq.~(\ref{del_1}) implies $\dot \Delta \simeq 0$ and assuming adiabatic initial conditions also $\Delta \simeq 0$. Thus in this limit the perturbations of DM and DR are locked to each other and described by Eq.~(\ref{didm_3}) with $\Delta = 0$. In this limit $\csp$ is the speed of sound of the locked IDM-DR fluid. Notice that as $\dN \rightarrow 0$ then $R \rightarrow \infty$ and $\csp^2 \rightarrow 0$ so that  \Eq{didm_3} reduces to that of $\delta_\cdm$ in \LC.

\subsection{Two limits}\label{subsec:lims}
\begin{figure}[!htbp]%
  \centering
  \includegraphics[width=0.8\textwidth]{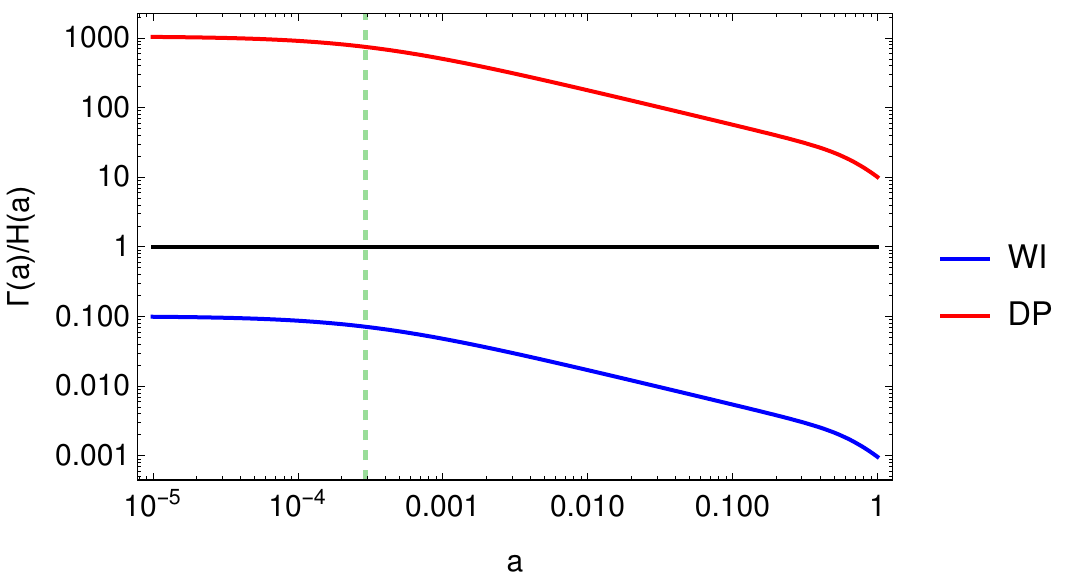}
  \caption{\label{fig:h_vs_g} Comparison of $\Gamma(a)$ and $H(a)$ for the limits WI and DP. Even though $\Gamma(a) \propto a^{-2}$, the ratio plotted has a changing slope because of the evolving $a$-dependence of $H(a)$. Note that for the DP limit it is sufficient to take $\g0 \gg \h0$, while for the WI we need $\Gamma \ll H$ during RD. The vertical dashed line is $a_\eq$, the scale factor at matter/radiation equality.}
\end{figure}

Despite the experimental tensions, \LCs does a fairly good job at describing the LSS data. Therefore we are mainly interested in small deviations from the \LCs predictions. This means that we will be mostly concerned with limits in which some of the parameters of the IDS are small. In the literature, two limiting cases of the IDS model have been recently studied:
\begin{itemize}
	\item \textbf{All DM is weakly interacting.} In this limit $f = 1$ and $\Gamma(a_\eq) \ll H(a_\eq)$ \cite{Buen-Abad:2015ova,Lesgourgues:2015wza}. As can be seen in \Fig{fig:h_vs_g}, this means that $\Gamma/H$ remains smaller than one during RD, and becomes even smaller at later times. This limit can be studied more easily with \Eqsto{didm_2}{ddr_2}. From now on we refer to this case as the \textit{Weakly Interacting} (\textit{WI}) limit. The WI model has the six free parameters of \LCs plus $\{\Gamma_0, \, \Delta N_\mathrm{fluid} \}$.
	\item \textbf{Only a fraction of the DM is IDM strongly coupled to DR}, $f \ll 1$, $\g0 \gg \h0$. This means that the IDM and the DR are tightly coupled today and, as shown in \Fig{fig:h_vs_g}, they have remained so since the early Universe. We then say that the IDM and DR together form a \textit{Dark Plasma} (\textit{DP}) \cite{Schmaltz,Chacko:2016kgg}. This scenario was dubbed \textit{``Partially Acoustic Dark Matter"} (\textit{PAcDM}) in \cite{Chacko:2016kgg}. The plasma has a speed of sound $\csp^2$ analogous to that of the baryon-photon plasma, given by the expression found in \Eq{cs2}. This limit can be understood more easily by studying \Eqsto{didm_3}{cs2}. As we shall see, in this limit, $\Gamma_0$ decouples from the leading order equations, thus the model has the six free parameters of \LCs plus $\{f, \, \Delta N_\mathrm{fluid} \}$.
\end{itemize}

We now briefly describe the behavior of the DM perturbations in these two limits of the IDS model. For a more detailed study see \App{appB}{B}.

\subsubsection*{Weakly Interacting}
In this limit all the DM is weakly interacting with the DR. From \Eqs{didm_2}{ddr_2} we can see that the friction term $\dot\delta_\idm$ gets a new contribution (apart from the usual Hubble expansion) coming from the momentum transfer rate $\Gamma$: $\mG \propto a^{-1}$. During the radiation dominated era $a \propto \eta^{-1}$ and therefore $\mH, \mG \propto \eta^{-1}$. This implies that those IDM modes that enter the Hubble radius during RD have a larger friction and thus a slower growth rate, i.e. these modes will be suppressed with respect to the \LCs case (see \Fig{fig:perts_l1}). Eventually, during MD $\mH \propto \eta^{-1}$ while $\mG \propto \eta^{-2}$. This means that the friction from the DR becomes negligible, and the equation for the IDM reduces to that of the CDM in \LC, with the solution $\delta_\idm \propto \eta^2$. For the same reason, modes that enter the Hubble radius after the friction from the DR has become irrelevant (sometime during MD) remain unsuppressed.

It is important to note that, because the IDM clumps less efficiently, the gravitational perturbations sourced by it are smaller.

\subsubsection*{Dark Plasma}
In this case the two fluids IDM and DR can be treated as a single one, obeying equation \Eq{didm_3} with $\Delta = 0$:
\begin{equation}\label{dp_eq}
	\ddot\delta_\idm + \frac{R}{1+R}\mH\dot\delta_\idm + k^2 \csp^2 \delta_\idm = -k^2 \psi + 3\ddot\phi + \frac{R}{1+R}3 \mH\dot\phi \ .
\end{equation}
This means that the IDM and DR perturbations track each other with $\delta_\dr = \frac{4}{3} \delta_\idm$,
as in the case of the tightly coupled baryon-photon plasma in $\LC$.

Early enough during RD $R \ll 1$ and thus $\csp^2 \approx 1/3$, which causes the modes inside the (dark) sound horizon to oscillate, as can be seen from \Eq{dp_eq}. This can be understood in terms of the pressure that the DR exerts on the IDM: because the two dark fluids are tightly coupled, the perturbations $\delta_\idm$ track the (oscillating) $\delta_\dr$, and thus do not grow nor form structure.

\begin{figure}[!htbp]%
	\centering
	\includegraphics[width=0.4\textwidth]{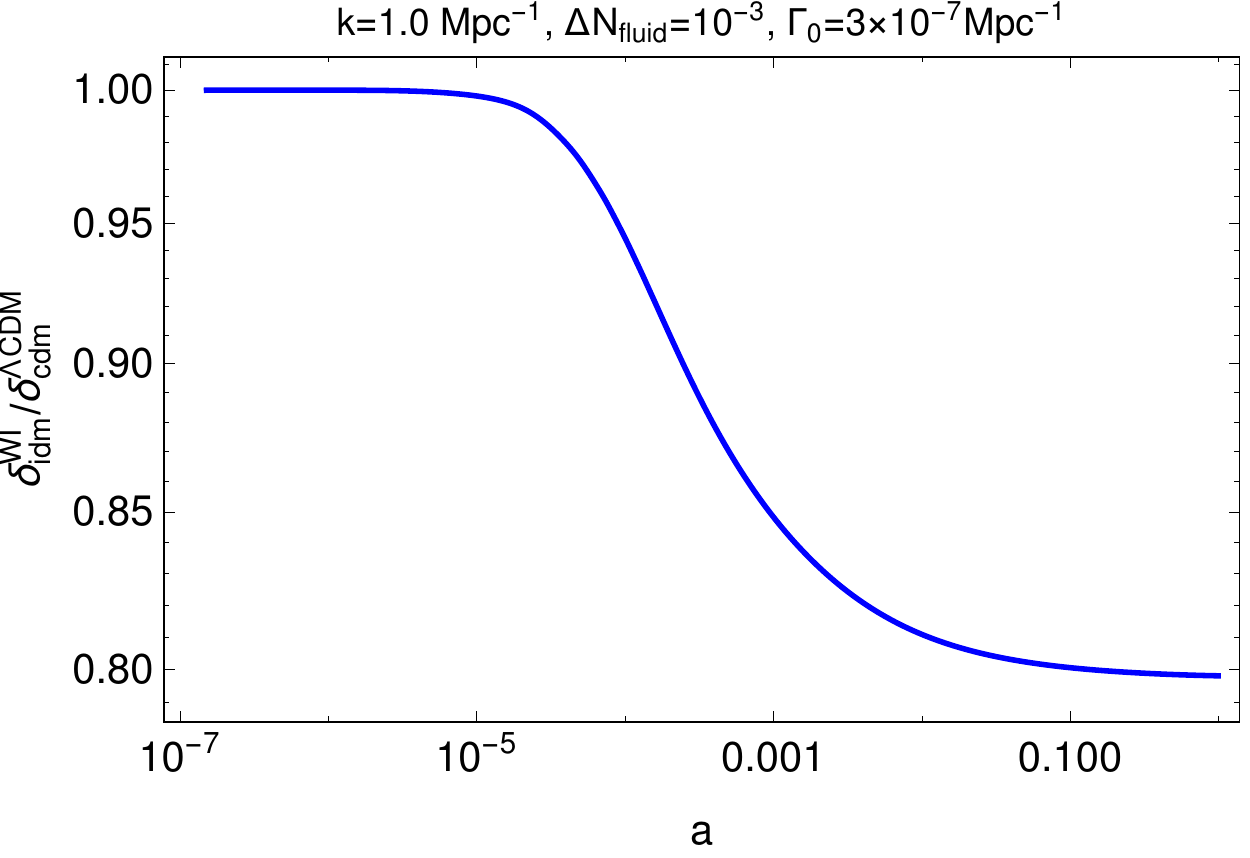}\label{fig:perts_l1}
	\includegraphics[width=0.52\textwidth]{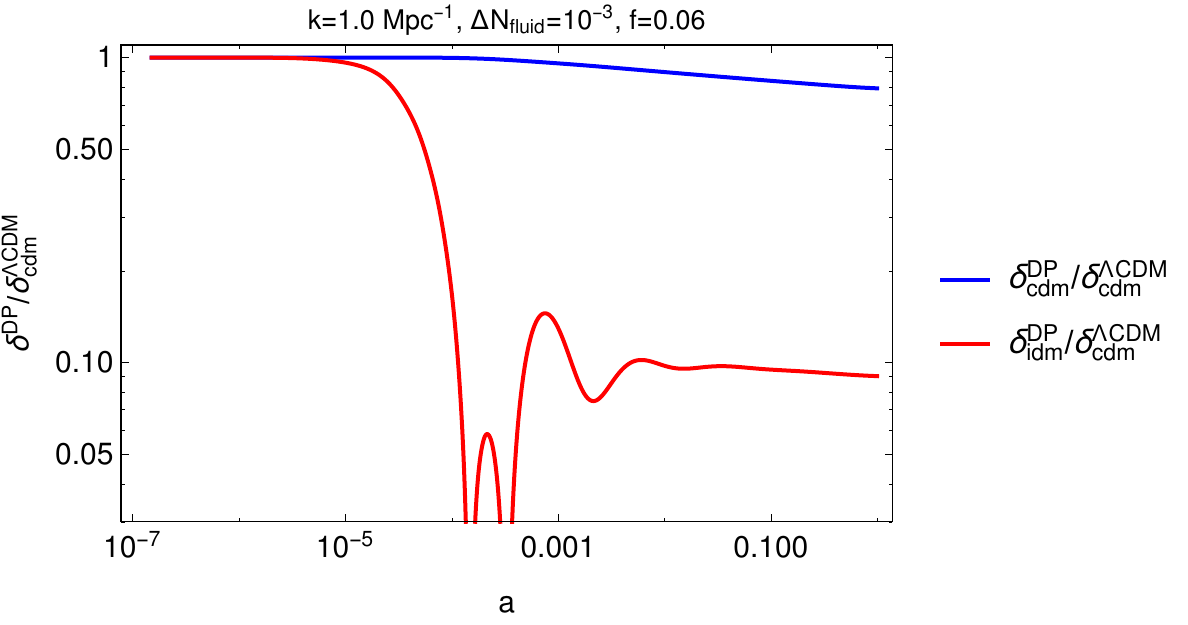}\label{fig:perts_l2}
	\caption{(\textit{Left}) Ratio of $\delta_\idm$ in WI to $\delta_\cdm$ in $\LC$. Note that at some point during MD the suppression saturates and remains more or less constant, because $\Gamma \propto a^{-2}$ decays faster than $H \propto a^{-3/2}$. (\textit{Right}) Ratio of $\delta_\cdm$ and $\delta_\idm$ in DP to $\delta_\cdm$ in $\LC$. Note that after horizon crossing these suppressions are never constant in time. Also, note that $\delta_\idm$ oscillates early on, but later has the same time dependence as $\delta_\cdm$ in DP: the two lines become parallel. The plots were made with CLASS (\cite{Blas:2011rf}), holding $\boldsymbol{\theta}_{\LC}$ and $\boldsymbol{\theta}_{\IDS}$ fixed.}
\end{figure}

The fraction $f$ of the DM that is IDM does not clump and therefore does not contribute to perturbations of the gravitational potential. The remaining $1-f$ fraction of standard CDM does source gravitational perturbations as usual, but the gravitational potential is now smaller by the factor of $1-f$. Thus the ordinary CDM sees a reduced gravitational potential and its density perturbations grow like $\delta_\cdm \sim \eta^{2-6 f/5}$ during MD; see \Eq{appB44} and its derivation, as well as \cite{Chacko:2016kgg}. Therefore even CDM perturbations grow slightly less than in \LCs, as shown in \Fig{fig:perts_l2}. 

The ratio $R$ keeps growing like the scale factor. Once $R>1$, the oscillations in $\delta_\idm$ are damped by the friction term in \Eq{dp_eq},
and the sound speed starts decreasing like $\csp^2 \propto a^{-1}$. Then the IDM perturbations start tracking the equilibrium solution given approximately by
$k^2 \csp^2 \delta_\idm = -k^2 \psi$. During MD and within the approximation $\Omega_\mathrm{b} \delta_\mathrm{b} \ll \Omega_\cdm \delta_\cdm$, the Poisson equation gives $-k^2 \psi \simeq 6 \eta^{-2} \Omega_\cdm \delta_\cdm$. Then the equilibrium solution reads $\delta_\idm = 6 (k \csp\eta)^{-2} \Omega_\cdm \delta_\cdm $, and since $\csp \propto \eta^{-1}$ the ratio between $\delta_\idm$ and
$\delta_\cdm$ becomes constant, as can be seen in  \Fig{fig:perts_l2}. Hence, for small wavelengths, $\delta_\idm$ remains much smaller than $\delta_\cdm$, and CDM fluctuations continue to grow at the slightly lower rate of $\delta_\cdm \sim \eta^{2-6 f/5}$ instead of the usual $\eta^2$ (\Eq{appB44}).
 
Note that this behavior is different from that of baryons and massive neutrinos, which behave as collisionless matter at late times (the former after the baryon drag epoch, the latter once their temperature decreases below their mass). Indeed, the speed of sound of both baryons and massive neutrinos scales like the ratio of their temperature and mass: $T/m \sim a^{-2} \sim \eta^{-4}$, which means that they cool down very fast and start falling into the gravitational potentials sourced by the CDM. On the other hand, in the DP model, the tight coupling between DR and IDM guarantees that the IDM temperature always tracks that of the DR, and the speed of sound prevents IDM perturbations to grow faster than and catch up to CDM perturbations. Thus $\delta_\idm$ and $\delta_\cdm$ do not reach a common value on small scales.

In the DP limit, the suppression of the DM perturbations once again translates into smaller gravitational perturbations sourced by them.

\section{Effects on the MPS, CMB spectrum, and CMB lensing}\label{sec:mps-cmb-lens}

The effects of a self-interacting DR fluid, governed by the parameter $\dN$, have already been described in several references. We will briefly recall these effects in the next paragraphs, assuming no IDM-DR interactions (i.e. $\Gamma_0=0$ or equivalently $f=0$). Then we will study the effects of the new parameters ($\Gamma_0$, $f$) in separate subsections.

The effect of $\dN$ on the observable LSS and CMB spectra can be decomposed into {\it background} and {\it perturbation} effects. The background effects are identical to those of extra free-streaming massless relics, usually parameterized by $\Delta N_\mathrm{eff}$. The perturbation effects are different for self-interacting and free-streaming degrees of freedom.

The major {\it background} effect of $\Delta N = \dN = \Delta N_\mathrm{eff}$ is best described by varying $\Delta N$ with a fixed redshift of radiation/matter and matter/$\Lambda$ equality (otherwise, the original effect of $\Delta N$ would be hidden by the trivial effect of a shift in these redshifts of equality)
\cite{Bashinsky:2003tk,Hou:2011ec,Lesgourgues:1519137}. This can be achieved by fixing four of the six \LCs parameters, namely $\{ \ob, \ n_\so, \ A_\so, \ \tau_\mathrm{reio} \}$, and varying the two remaining ones $\{\theta_\so, \odmtot\}$ plus $\Delta N$ in such a way that the total density of radiation, matter and cosmological constant get rescaled by the same number. Hence the critical density today is enhanced, and the Hubble parameter $H_0$ (or the reduced Hubble parameter $h$) must increase. Under this transformation, the three characteristic distances playing a role in the CMB spectra, which are the angular diameter distance to decoupling, the sound horizon at decoupling and the diffusion damping scale at decoupling, evolve respectively like $d_A (z_\mathrm{dec}) \propto h^{-1}$, $d_\so(z_\mathrm{dec}) \propto h^{-1}$ and $d_\mathrm{d}(z_\mathrm{dec}) \propto h^{-1/2}$. Then the angle of the peaks given by $\theta_\so = d_\so / d_A$ is preserved, but the angle of the Silk damping envelope $\theta_\mathrm{d} = d_\mathrm{d}/d_A$ is not. Hence the main background effect of varying $\Delta N$ is to change the ratio between the Silk damping angular scale and the acoustic peak angular scale. 

The {\it perturbation} effects of $\dN$ are much smaller than those of an equivalent $\Delta N_\mathrm{eff}$ (see e.g. \cite{Audren:2014lsa} and references therein, or \cite{Lesgourgues:2015wza,Oldengott:2017fhy}). Extra free-streaming massless particles travel at the speed of light $c=1$ and pull the CMB peaks towards larger scales (smaller angles) through a neutrino drag effect~\cite{Bashinsky:2003tk,Hou:2011ec,Lesgourgues:1519137}. Instead, self-interacting DR features acoustic oscillations propagating at a sound speed $c_\dr^2=1/3$ (or $0<\csp^2<1/3$ for a tightly-coupled IDM-DR fluid) and do not produce such an effect. Besides, the CMB spectrum is sensitive to the gravitational interactions between photon perturbations and extra relic perturbations before decoupling. In the case of extra free-streaming massless particles, photons
couple with a very smooth component, and the CMB spectrum amplitude is slightly reduced on scales crossing the sound horizon before decoupling \cite{Bashinsky:2003tk,Hou:2011ec,Lesgourgues:1519137}. In the case
of a self-interacting fluid, the photon fluid couples with a DR fluid with a comparable fluctuation amplitude, thus no such suppression is observed.

Overall, the effect of $\dN$ on the CMB is smaller than that of an equivalent $\Delta N_\mathrm{eff}$, leading to weaker bounds. Instead, the effects of
$\dN$ or $\Delta N_\mathrm{eff}$ on the MPS are roughly equal, because they are both dominated by {\it background} effects. Assuming the same transformation as before, which is such that $\Delta N$ increases while $\{ \omega_\mathrm{b}, \ z_\mathrm{eq}, \  \Omega_\Lambda\}$ are constant, we find that the ratio
$\omega_\mathrm{b}/\omega_\mathrm{cdm}$ must vary. This ratio affects the small-scale amplitude of the MPS. Models with larger $\Delta N$ should have
 a smaller ratio $\omega_\mathrm{b}/\omega_\mathrm{cdm}$ and a thus higher MPS amplitude on small wavelengths/large wavenumbers~\cite{Lesgourgues:1519137}, as well as a higher CMB lensing spectrum amplitude on small angles/large multipoles. 

We now turn to the description of the effects of the IDM-DR interaction, governed by $\g0$ in the WI model, and by $f$ in the DP model. Throughout the next subsections, we hold the \LCs parameters $\boldsymbol{\theta}_{\LC}$ fixed to their best fit values in \cite{Ade:2015xua}, and $\dN$ fixed arbitrarily to 0.4. We compare the LSS and CMB spectra obtained with growing values of $\g0$ or $f$ to a reference \LC$+\dN$ model with the same $\dN=0.4$. In each of the next subsections, we will review the effects of $\g0$ or $f$ on, respectively, the MPS, the CMB lensing spectrum, and the CMB temperature spectrum.

\subsection{Matter Power Spectrum}\label{subsec:mps}
\subsubsection*{Weakly Interacting}

For $\g0 > 0$, the effect of the friction on the IDM perturbations with $k \gg k_\eq$, discussed in \Sec{sec:ids}, translates into a suppression in the MPS as observed today, shown in \Fig{fig:mps_l1} (see also \cite{Buen-Abad:2015ova,Lesgourgues:2015wza}). What is interesting is that the suppression in the MPS is $k$-dependent: the larger wavenumbers were inside the Hubble radius (and thus felt the friction from DR) during RD for longer. Hence this suppression is not step-shaped like for massive neutrinos. Roughly speaking, it would resemble a step in the effective spectral index of the MPS, with a lower index  for $k\geq k_\mathrm{\eq}$. More precisely, in the limit of small $\g0$ and for $k > k_\eq$, the suppression factor is $\sim (1 -\frac{\sqrt{2}\mG}{\mH}\big\vert_\eq \log k \eta_\eq)$; see \Eq{appB30} and its derivation.

\subsubsection*{Dark Plasma}
Let us now consider the effect of $f$ on the MPS of the DP model, and compare it to $\LC+\dN$ with the same $\dN$. As mentioned before (and posited originally in \cite{Schmaltz,Chacko:2016kgg}), the fraction of DM that is IDM is so strongly coupled to the DR that the $\delta_\idm$ perturbations, in their tracking of $\delta_\dr$, oscillate and are therefore temporarily prevented from clumping and growing. This happens only on sub-Hubble scales and as long as the speed of sound $\csp^2$ is sizable: hence, only small wavelengths with typically $k \gg k_\eq$ experience this regime. For these scales, once $\csp^2$ becomes sufficiently small, the $\delta_\idm$ perturbations stop oscillating, but remain smaller than $\delta_\cdm$. Hence the MPS is suppressed on small scales for two reasons: the negligible contribution of $\delta_\idm$ to the total matter fluctuations, bringing a factor $(1-2f)$, and the reduced growth rate of $\delta_\cdm \sim \eta^{2-6f/5}$, bringing approximately a factor $\bl( \frac{\eta_0}{\eta_\eq} \br)^{-12f/5}$. In total the small-scale MPS is suppressed by $\sim (1-2f)\bl( \frac{\eta_0}{\eta_\eq} \br)^{-12f/5}$ (\Eq{appB46}). A detailed derivation of this suppression can be found in our \App{appB}{B} and in \cite{Chacko:2016kgg}.

\begin{figure}[!htbp]%
    \centering
    \includegraphics[width=0.45\textwidth]{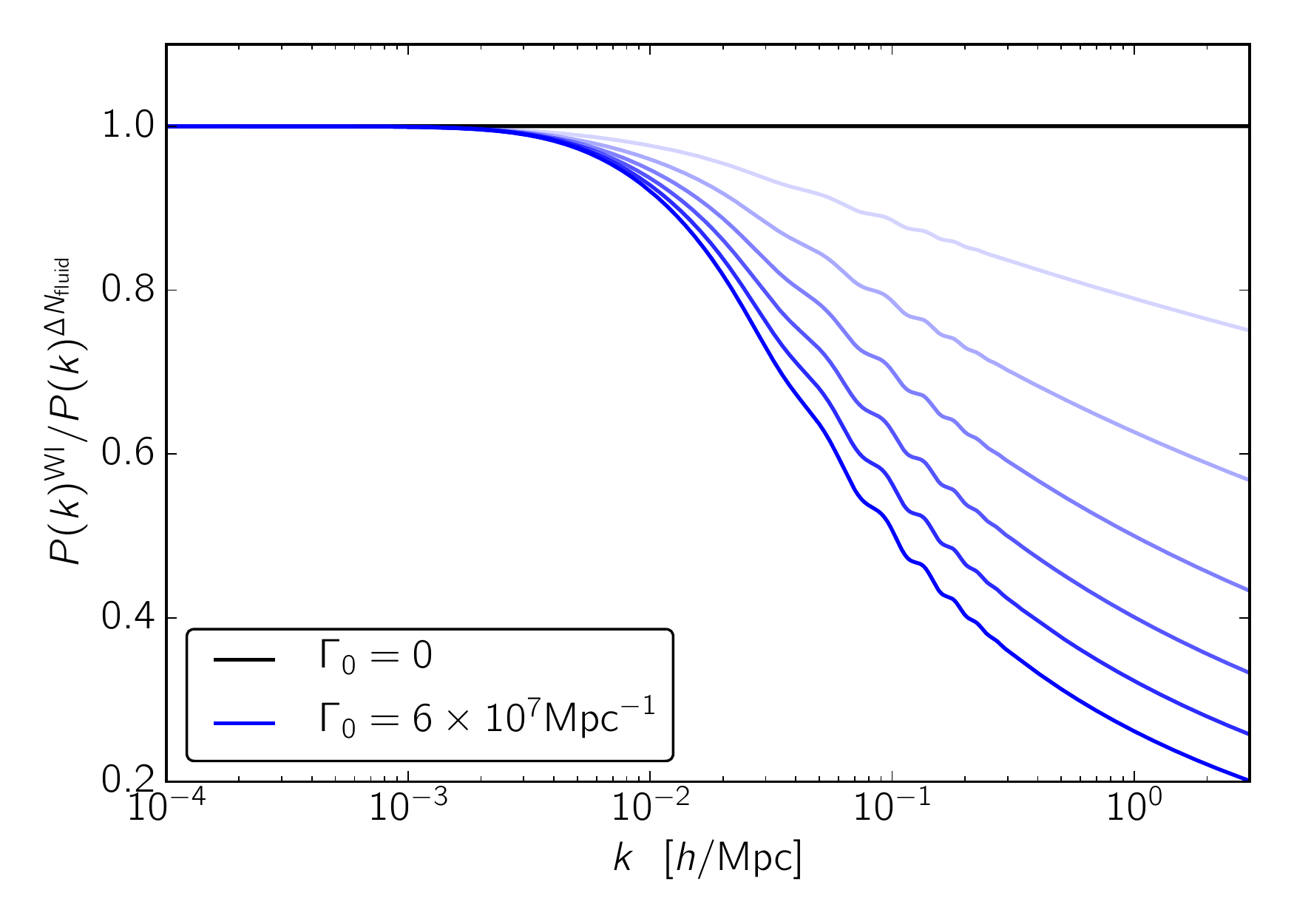}\label{fig:mps_l1}
    \includegraphics[width=0.46\textwidth]{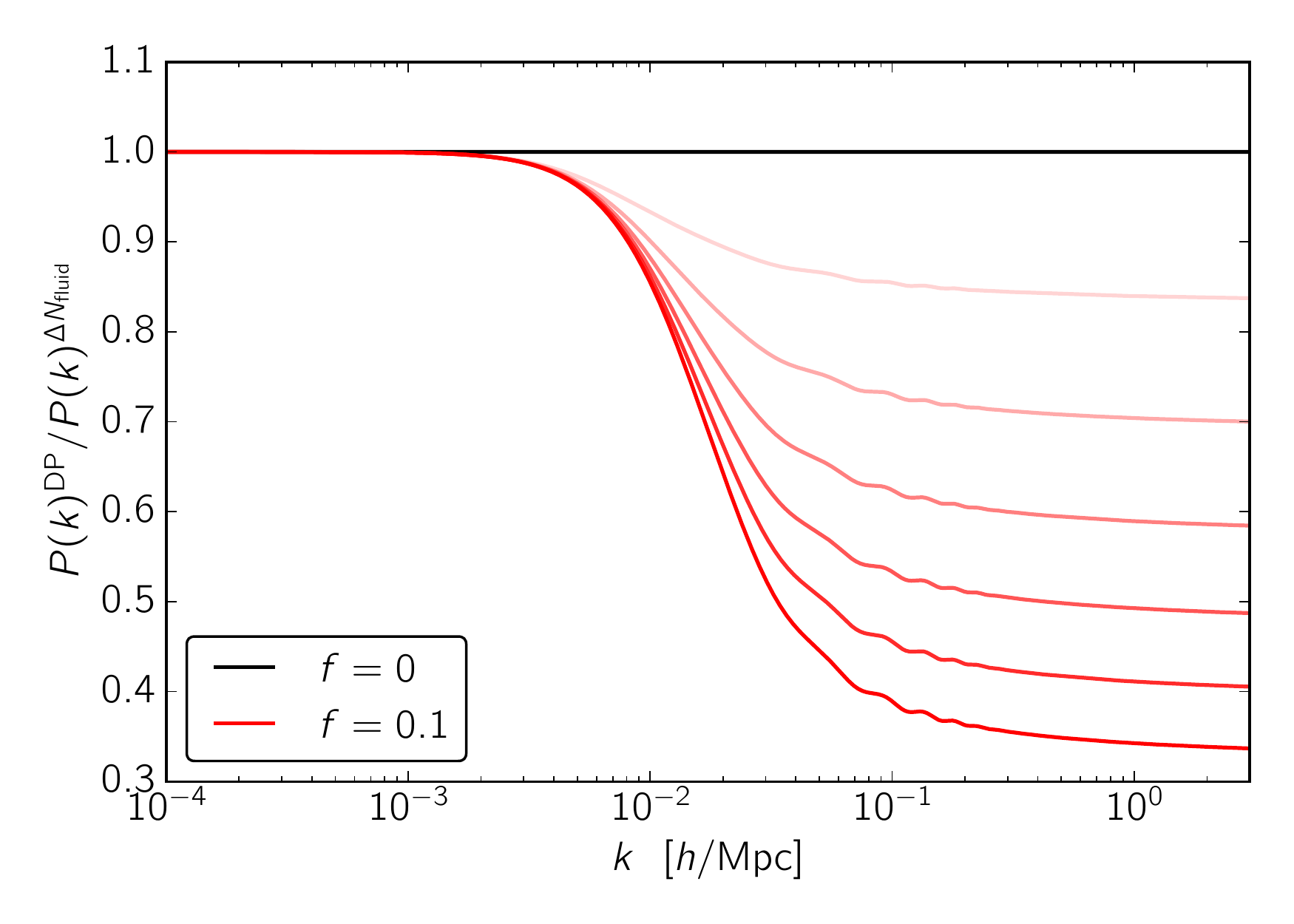}\label{fig:mps_l2}
    \caption{CLASS plots of the ratio of the linear MPS from the IDS model to that from $\LC+\dN$ (\textit{left}) in the WI limit, for different $\g0$; and (\textit{right}) in the DP limit, for different $f$. Note the $k$ (in)dependence of the suppression in the left (right) plots.}
\end{figure}

This effect is qualitatively similar to that of massive neutrinos, and also leads to a step-like suppression of the MPS compared to that of the equivalent $\LC+\dN$ model, as can be seen in \Fig{fig:mps_l2}. However, the characteristic times and scales involved in our model are different. In the massive neutrino model, the step in the linear MPS is located at a scale $k_\mathrm{nr}$ and has an amplitude $(1-8f_\nu)$, where ($k_\mathrm{nr}$, $f_\nu$) are given respectively by the individual and total neutrino masses~(see e.g. \cite{Hu:1997mj,Lesgourgues:2006nd,Lesgourgues:1519137}). In the DP model, the scale of the step is $k_\eq$ with an amplitude of the suppression as given in the previous paragraph.

\subsection{CMB lensing}\label{subsec:lens}

The CMB lensing potential $C_\ell^{\phi\phi}$ is given in the Limber approximation (\cite{Limber:1954zz,Pan:2014xua}) by:
\begin{equation}\label{limber}
	\ell^4 C_\ell^{\phi\phi} \approx 2 \int\limits_{0}^{\chi_\dec}d\chi \bl( \frac{\ell}{\chi} \br)^4 P_{(\phi+\psi)} \bl( k=\frac{\ell}{\chi} ; a(\chi) \br) \bl( 1 - \frac{\chi}{\chi_\dec} \br)^2 \ ,
\end{equation}
where $\chi$ is the comoving distance as measured from the observer, and $P_{(\phi+\psi)} $ is the Power Spectrum of the sum of the metric perturbations, related to that of matter fluctuations on sub-Hubble scales by the Poisson equation. Hence the impact of different cosmological model on the MPS and CMB lensing spectrum is almost identical.

\begin{figure}[!htbp]%
    \centering
    \includegraphics[width=0.45\textwidth]{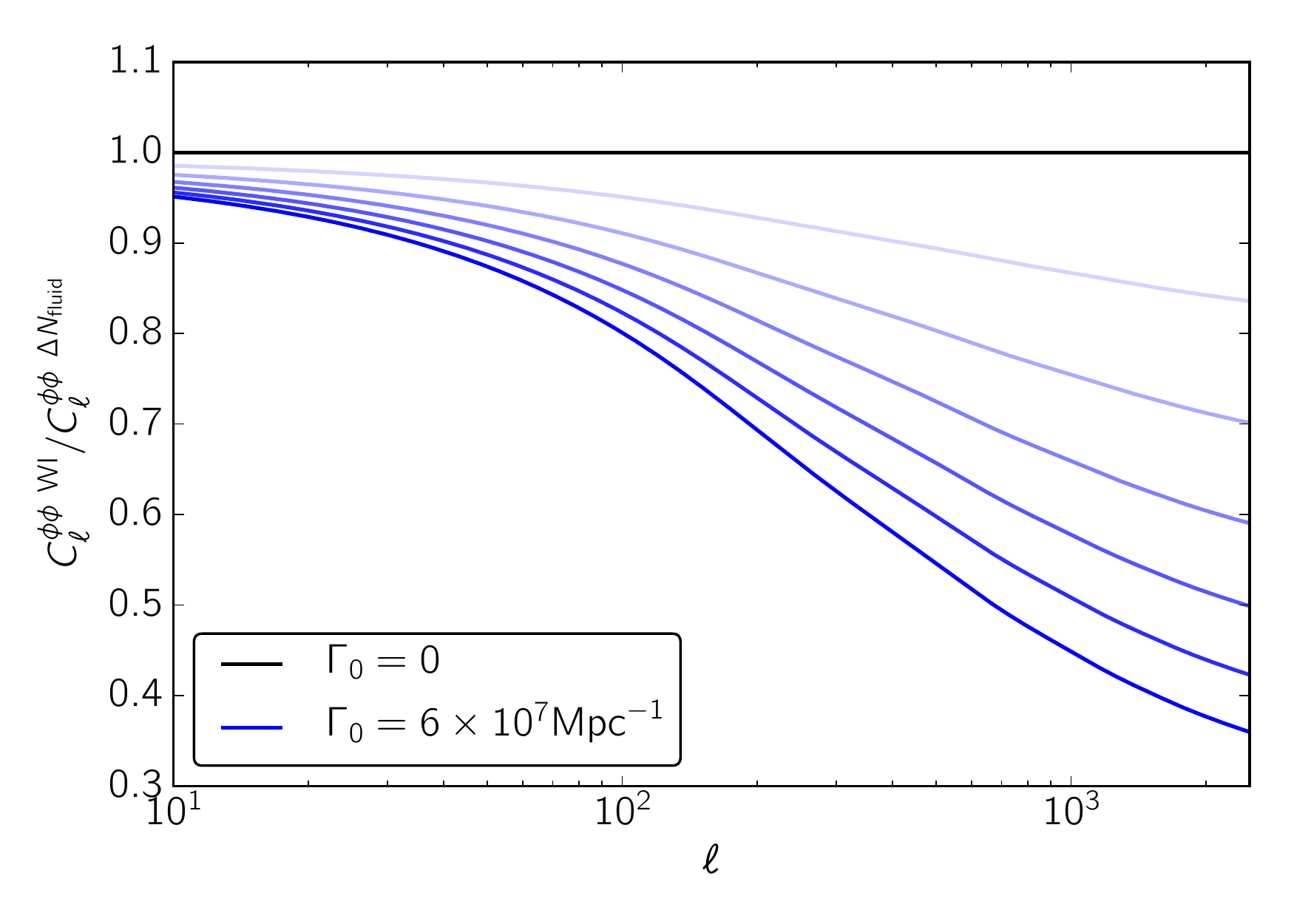}\label{fig:lens_g0}
    \includegraphics[width=0.45\textwidth]{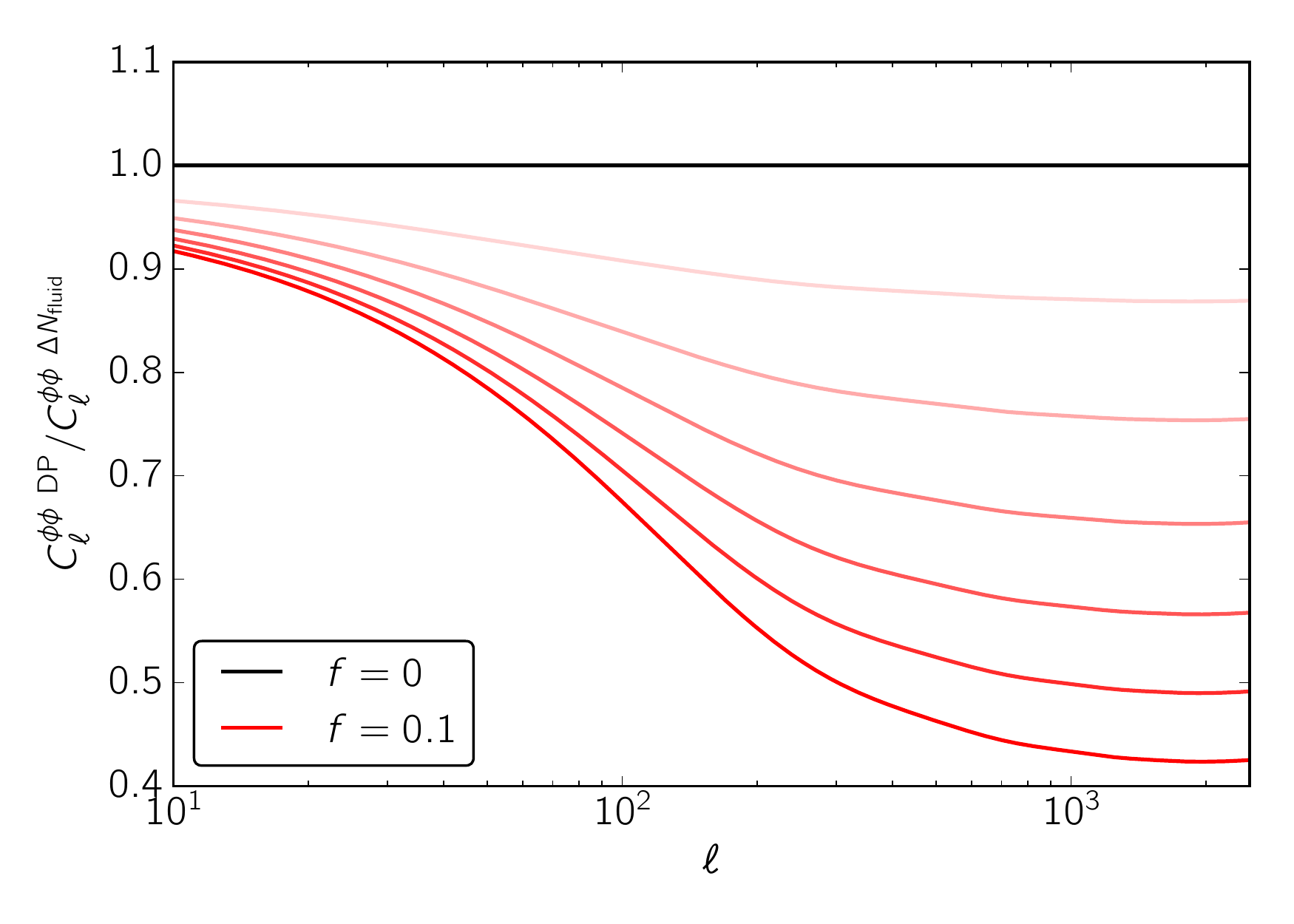}\label{fig:lens_f}
    \caption{CLASS plots of the ratio of the CMB lensing spectrum from the IDS model to that from $\LC+\dN$ (\textit{left}) in the WI limit, for different $\g0$; and (\textit{right}) in the DP limit, for different $f$.}%
\end{figure}

\Fig{fig:lens_g0} and \Fig{fig:lens_f} show the effects of $\g0$ and $f$ on the lensing power spectrum $C_\ell^{\phi \phi}$. These two parameters produce a smaller lensing spectrum due to the suppression in the DM perturbations yielding shallower gravitational perturbations.

\subsection{CMB spectrum}\label{subsec:cmb}
\subsubsection*{Weakly Interacting}

\begin{figure}[!htbp]%
\centering
 \includegraphics[width=0.90\textwidth]{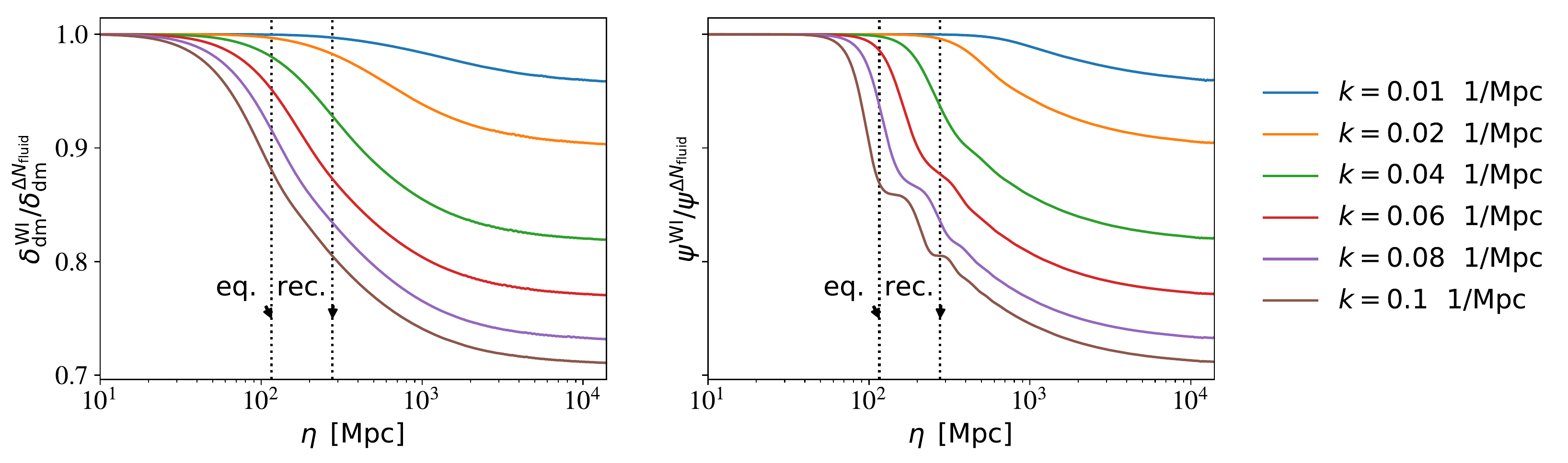}
    \caption{CLASS plots of the ratio of perturbations $\delta_\mathrm{dm}(\eta)$ and $\psi(\eta)$ from the WI model with $\Gamma_0 = 6\times 10^{-7} \mathrm{Mpc}^{-1}$ to that from the $\Lambda$CDM+$\Delta N_\mathrm{fluid}$ model, both with $\Delta N_\mathrm{fluid}=0.4$, for several wavenumbers $k$ relevant for the first CMB peaks. The vertical lines show the conformal time at radiation/matter equality and at recombination, and the maximum value of $\eta$ corresponds to the conformal time today. \label{fig:delta_psi_wi}}%
\end{figure}

The effect of the DM-DR interaction on the CMB spectra is a little bit more subtle than that on the matter power spectrum. The final effect does not depend directly on the perturbation $\delta_\mathrm{dm}(k, \eta)$ anymore, but rather on the metric fluctuations $\phi(k, \eta)$ and $\psi(k, \eta)$.
The left plot in \Fig{fig:delta_psi_wi} shows how $\delta_\mathrm{dm}(\eta,k)$ is suppressed for various wavenumbers due to the DM-DR interaction.
The metric perturbation $(\phi, \psi)$ have a similar behavior, although the suppression starts at a later time for each mode. The reason is that the metric perturbations track the non-relativistic matter perturbations (of IDM plus baryons) only when the modes are deep inside the Hubble radius.

This enhanced damping of metric fluctuations has non-trivial implications on the CMB temperature spectrum, both before recombination (through the intrinsic temperature and Sachs-Wolfe term $[\delta_\gamma/4+\psi]$) and soon after recombination (through the early Integrated Sachs-Wolfe (ISW) effect). The effects of $\g0$ on the unlensed CMB temperature spectrum is shown in \Fig{fig:tt_g0} (left plot). A detailed study of the behavior of the perturbations shows that the different time evolution of the metric fluctuations changes the amplitude and the zero-point of the oscillations of the variable $[\delta_\gamma/4+\psi]$, in such a way that with a higher $\Gamma_0$, the first acoustic peak is slightly enhanced, while all other peaks are suppressed. In addition, the early ISW contribution to $C_\ell^{TT}$ is shifted to higher multipoles, further contributing to the enhancement of the first peak, and raising the spectrum between the first peak and the first dip.

On top of these effects, the  observed CMB spectrum is affected by CMB lensing. The reduction of amplitude of $C_\ell^{\phi\phi}$ discussed in section~\ref{subsec:lens} implies that for a higher $\Gamma_0$, the observable CMB spectrum is slightly less affected by lensing, showing therefore more contrast between maxima and minima.

\subsubsection*{Dark Plasma}
\begin{figure}[!htbp]%
    \centering
    \includegraphics[width=0.45\textwidth]{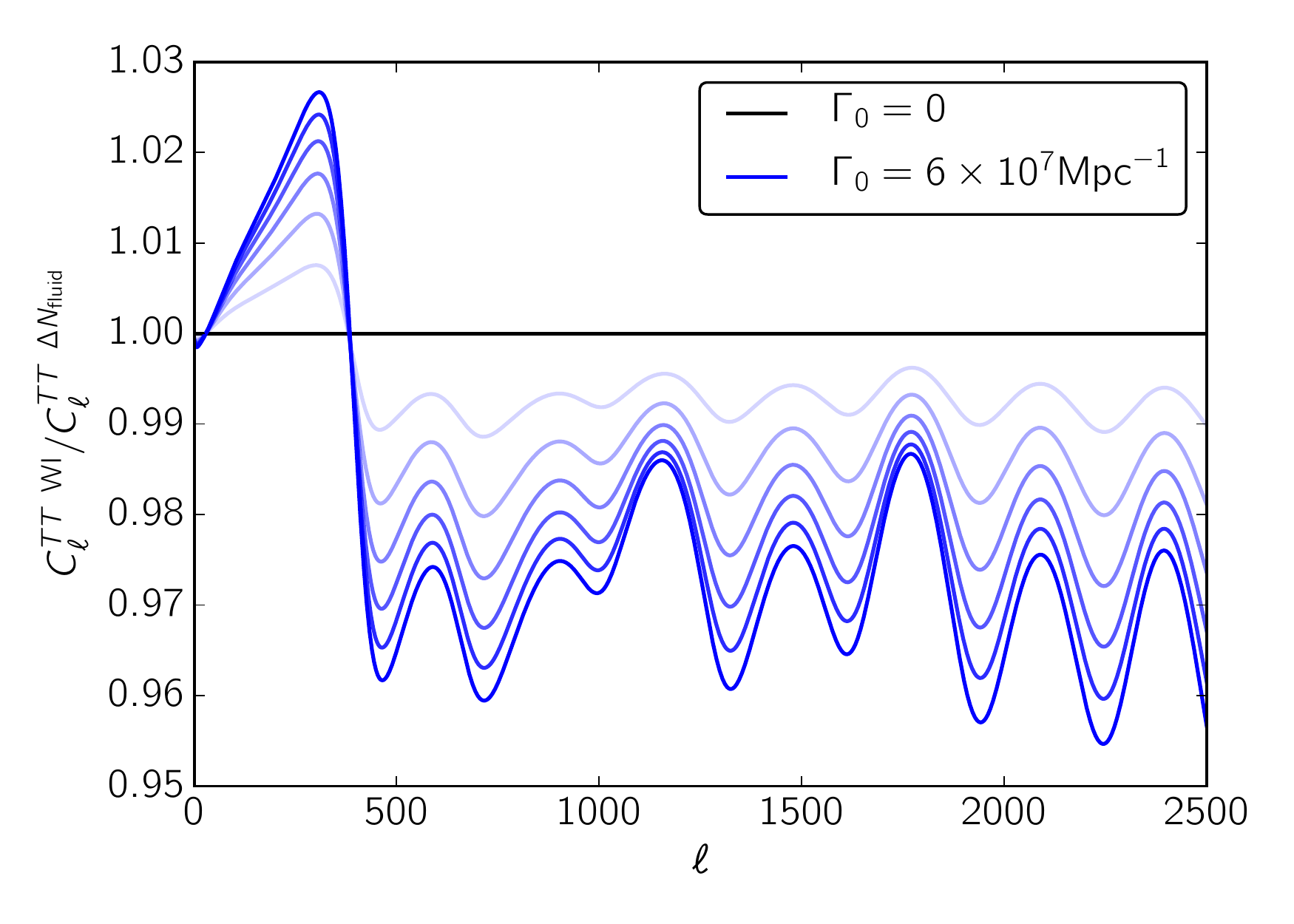}\label{fig:tt_g0}
    \includegraphics[width=0.45\textwidth]{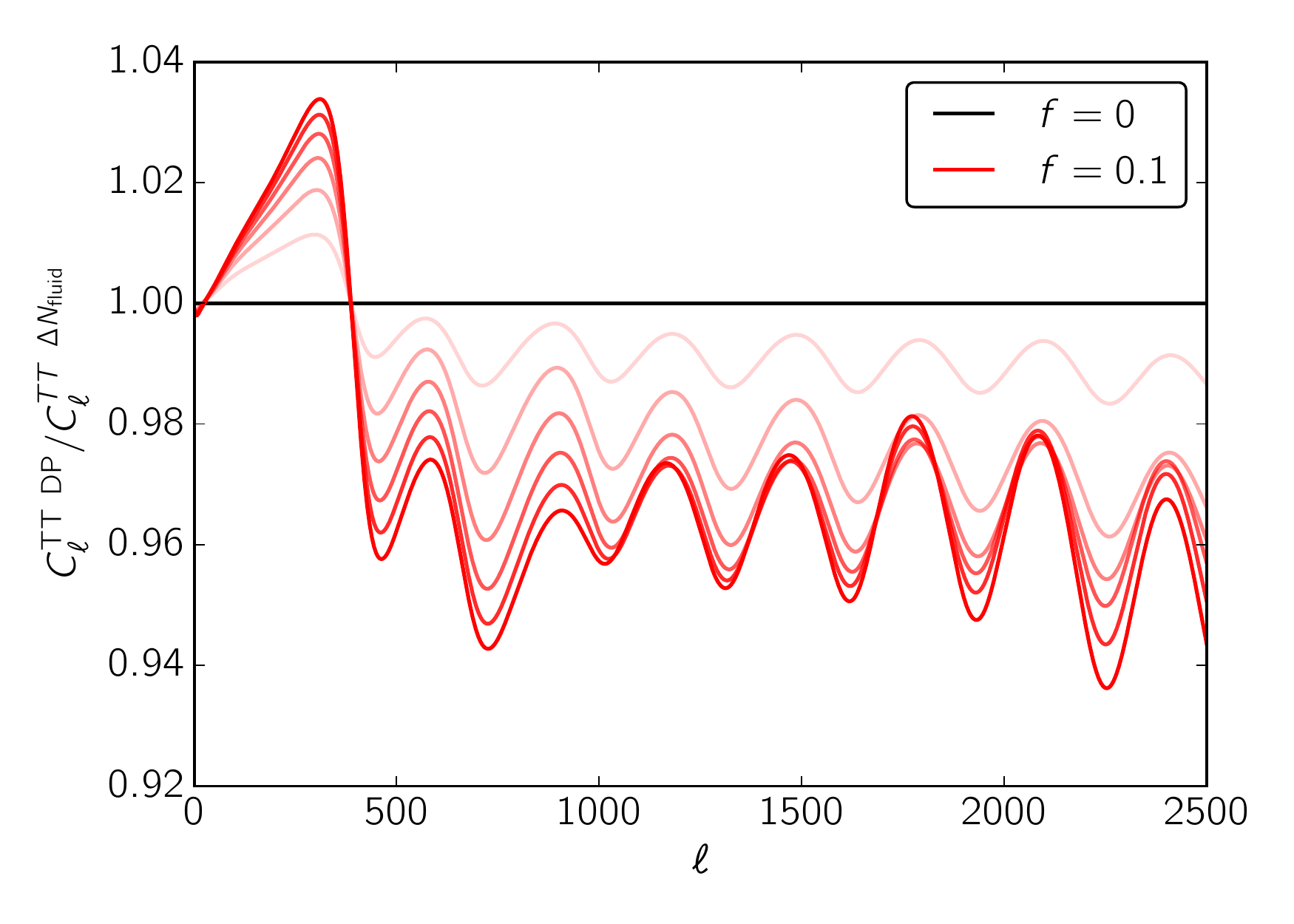}\label{fig:tt_f}
    \caption{CLASS plots of the ratio of the unlensed temperature spectrum from the IDS model to that from $\LC+\dN$ (\textit{left}) in the WI limit, for different $\g0$; and (\textit{right}) in the DP limit, for different $f$. Note the reduction in the suppression for larger $f$ and high multipoles, due to contributions to the Integrated Sachs-Wolfe Effect during the matter dominated era.}
\end{figure}

\Fig{fig:tt_f} shows the effects of $f$ on the unlensed TT spectrum of the DP model. As in the WI case, non-zero $f$ means that the time evolution of the gravitational perturbations is modified, and therefore so is the Early Integrated Sachs-Wolfe Effect around the first peak. Also, for larger $\ell$, the suppression of the gravitational perturbations due to the reduced clumping rate of the CDM leads to a reduction of the $C_\ell$ for $\ell \geq 400$.

Finally, notice the curious behavior of the spectrum suppression for larger and larger $f$: the suppression is actually reduced compared to that for smaller $f$. This is due to the fact that during MD the gravitational potentials in the DP model do not remain constant like they do in $\LC+\dN$ (or for that matter $\LC$), but have an exponential dependence on $f$ ($k^2 \psi \sim \eta^{-6f/5}$). This means that the Integrated Sachs-Wolfe contribution to $C_\ell^{TT}$ has an extra contribution from the Matter Dominated era, thus enhancing the spectrum.

\newpage

\section{Results}\label{sec:results}

We implemented the IDS model into the Boltzmann code CLASS (\cite{Blas:2011rf}) and use MontePython \cite{Audren:2012wb}, (in some cases with MultiNest \cite{Feroz:2007kg,Feroz:2008xx,Feroz:2013hea,Buchner:2014nha}), to fit to experimental data currently available and to produce the plots in this section. We run with three massive neutrinos, with $m_\nu = 0.02 \ \mathrm{eV}$ each (since current data is mainly sensitive to the total neutrino mass; this is known to be a good enough approximation to the Minimal Normal Hierarchy scenario). Only minor modifications to the CLASS code are necessary in order to include the IDM and DR. The theoretically motivated regime of $\dN \geq 0.07$ (see \cite{Buen-Abad:2015ova}) was explored in \cite{Lesgourgues:2015wza} for the WI limit. In this work, we repeat the analysis of the WI limit and fit to newer data, and we also do this for the DP limit. Finally, we also explore the small $\dN$ regime through a flat prior on $\log_{10} \dN$.

In summary, we have six different cases to which we fit the data: $\LC$; WI and DP limits with, for each of them, either a linear prior $\dN \geq 0.07$ or a log prior $-5 \leq \log_{10} \dN \leq 0$; and the general IDS model, with log priors on the three parameters $\boldsymbol{\theta}_{\IDS}$, which in this case are allowed to float.

\subsection{The Experiments}\label{subsec:exps}

We divide the data into the following sets:
\begin{itemize}
	\item \textbf{CMB:} For high multipoles, we use the Planck 2015 high-$\ell$ TT+TE+EE data (\cite{Aghanim:2015xee}). Besides, some recent intermediate Planck results removed previously unaccounted for systematics in the low-$\ell$ region of the polarization spectra, and produced a gaussian posterior distribution for $\tau_\mathrm{reio} = 0.055 \pm 0.009$ (see \cite{Aghanim:2016yuo,Adam:2016hgk}). Because the improved low-$\ell$ data is not publicly available at the time of writing of this paper, we use this $\tau_\mathrm{reio}$ posterior instead of the Planck 2015 low-$\ell$ likelihood.
	\item \textbf{BAO:} We use measurements of $D_V/r_\mathrm{drag}$ by 6dFGS at $z = 0.106$ (\cite{Beutler:2011hx}), by SDSS from the MGS galaxy sample at $z = 0.15$ (\cite{Ross:2014qpa}), and by BOSS from the CMASS and LOWZ galaxy samples of SDSS-III DR12 at $z = 0.2-0.75$ (\cite{Alam:2016hwk}).
	\item \textbf{LSS:} We use the following Large Scale Structure information: the Planck 2015 gravitational lensing likelihood (\cite{Ade:2015zua}), the $\sigma_8 (\Omega_\ma / 0.27)^{0.30} = 0.782 \pm 0.010$ ($68 \%$ C. L.) constraint from Planck SZ cluster counts (\cite{Ade:2013lmv}), the full correlation functions measured by the CFHTLens weak lensing survey~\cite{Heymans:2013fya} (after using the updated version of HALOFIT \cite{Takahashi:2012em} to treat the non-linearities of the MPS, see \cite{MacCrann:2014wfa}), and the measurement of the halo power spectrum from the Luminous Red Galaxies SDSS-DR7 (\cite{Reid:2009xm}).
	\item $\mathbf{H_0}$\textbf{:} We also include the latest result on the direct measurement of the Hubble parameter by Adam Riess et al., $\h0 = 73.24 \pm 1.74 \ \mathrm{km\ s^{-1}\ Mpc^{-1}}$ (\cite{Riess:2016jrr}).
\end{itemize}

It is usually hazardous to combine data sets contradicting each other. In our case, direct measurements of $H_0$ or constraints on $\sigma_8$ from CFHTLens and Planck SZ clusters are known to be in tension with other datasets in the framework of $\Lambda$CDM. This is not the same as saying that the data sets contradict each other. The values of $H_0$ or $\sigma_8$ inferred from Planck are not directly measured, they are just extrapolated from the best-fitting model in the particular framework of, e.g., a $\Lambda$CDM cosmology. Since HST, CFHTLens, Planck SZ clusters and other data sets probe different quantities, they are not in direct contradiction. The actual important relevant question is to find whether they can be brought in good agreement with each other in the context of an extended cosmological scenario. Hence it is perfectly legitimate to combine all these data sets together in the context of IDS models. Our goal is to check whether the best-fitting extended model is a reasonable fit of each individual data set, in which case some positive conclusions could be drawn;
or the result of a compromise between data sets still being in tension with each other, in which case we would need to remain very careful concerning the final interpretation.

\subsection{Numerical Results}\label{subsec:nums}

\subsubsection{The best fit $\chi^2$}

In table~\ref{tab:chi2} we show the minimum value of $\chi^2_\mathrm{eff} =-2\ln {\cal L}$ for each run. The most striking result is the amount by which this number gets reduced with the IDS model. This is especially true when we allow for very small values of $\dN$, covered by the logarithmic prior $-5 \leq \log_{10} \dN \leq 0$. In that case, we obtain $\Delta \chi^2_\mathrm{eff} \simeq -22.2$ ($-20.0$) with just two extra free parameters in the WI (DP) DP limit; or $\Delta \chi^2_\mathrm{eff} \simeq -23.7$ with three extra free parameters $\boldsymbol{\theta}_{\IDS}$ in the general case. Since \LCs is contained in the larger parameter spaces of both the WI and DP models we can quote a significance at which the best fit regions are preferred over the \LCs fit.
We find a $4.3\sigma$ ($4.1\sigma$) preference for the best fit point of the WI (DP) extended model, or $4.2\sigma$ for that of the general IDS model. In the case of the linear prior with the restriction $\dN \geq 0.07$ motivated by some classes of IDS models, we do not cover the best-fit region of parameter space with very small DR densities. In that case the preference for the IDS model is still there but less significant, with: $\Delta \chi^2_\mathrm{eff} \simeq -14.3$ with two free parameters in the WI limit ($\sim 3.4\sigma$ preference), and $\Delta \chi^2_\mathrm{eff} \simeq -10.3$ with two free parameters in the DP limit ($\sim 2.8\sigma$ preference).

\begin{table}
	\begin{tabular}{|| c || c | c | c | c | c ||} 
		\hline
		\multicolumn{6}{|c|}{The best-fit $\chi^2$ per experiment of each model} \\
		\hline\hline
		\multirow{2}{*}{Data Sets} & \multirow{2}{*}{\LC} & \multicolumn{2}{c |}{WI limit} & \multicolumn{2}{c |}{DP limit} \\ [0.5ex] 
		 & & \tiny{$\dN$ $\log$ Prior} & \tiny{$\dN$ lin. Prior} & \tiny{$\dN$ $\log$ Prior} & \tiny{$\dN$ lin. Prior} \\
		\hline\hline
		high-$\ell$ TTTEEE & $2452.6$ & $2446.03$ & $2455.22$ & $2447.50$ & $2450.54$   \\
		\hline
		SimLow $\tau_\mathrm{reio}$ & $0.34$ & $0.03$ & $0.07$ & $0.12$ & $0.67$   \\
		\hline
		BAO & $15.33$ & $13.69$ & $13.45$ & $13.50$ & $14.21$   \\
		\hline
		lensing & $10.43$ & $9.53$ & $11.50$ & $9.35$ & $10.34$   \\
		\hline
		SDSS & $45.43$ & $45.06$ & $45.83$ & $44.08$ & $45.56$  \\
		\hline
		CFHTLens & $100.00$ & $100.41$ & $98.41$ & $101.46$ & $98.76$  \\
		\hline
		Planck SZ & $15.50$ & $0.05$ & $3.62$ & $0.52$ & $7.6$   \\
		\hline
		$\h0$ & $7.80$ & $9.44$ & $4.00$ & $8.93$ & $8.39$   \\
		\hline\hline
		TOTAL & $2646.42$ & $2624.23$ & $2632.09$ & $2626.47$ & $2636.08$  \\
		\hline
		$\Delta \chi^2_\mathrm{eff}$ & 0 & $-22.19$ & $-14.33$ & $-19.95$ & $-10.34$   \\ [1ex] 
		\hline\hline
	\end{tabular}
	\caption{Minimum {\it effective chi square} $\chi^2_\mathrm{eff} =-2\ln {\cal L}$ for the Weakly Interacting and Dark Plasma model, with the contribution from each individual data set.\label{tab:chi2}}
\end{table}

\begin{table}
	\begin{tabular}{|| c || c | c ||} 
		\hline
		\multicolumn{3}{|c|}{The best-fit $\chi^2$ per experiment of each model} \\
		\hline\hline
		\multirow{2}{*}{Data Sets} & \multirow{2}{*}{\LC} & General IDS \\ [0.5ex] 
		& &  $\boldsymbol{\theta}_\mathrm{IDS}$ $\log$ Prior \\
		\hline\hline
		TTTEEE lite & $575.10$ & $567.78$ \\
		\hline
		SimLow $\tau_\mathrm{reio}$ & $0.26$ & $0.051$ \\
		\hline
		BAO & $16.48$ & $13.40$ \\
		\hline
		lensing & $10.13$ & $9.35$ \\
		\hline
		SDSS & $45.77$ & $44.02$ \\
		\hline
		CFHTLens & $98.56$ & $99.78$ \\
		\hline
		Planck SZ & $13.68$ & $0.19$ \\
		\hline
		$\h0$ & $7.00$ & $8.74$ \\
		\hline\hline
		TOTAL & $766.98$ & $743.32$ \\
		\hline
		$\Delta \chi^2_\mathrm{eff}$ & $0$ & $-23.66$ \\ [1ex] 
		\hline\hline
	\end{tabular}
	\caption{Minimum {\it effective chi square} $\chi^2_\mathrm{eff} =-2\ln {\cal L}$ for the general Interacting Dark Sector model, with the contribution from each individual data set.\label{tab:chi2_gen}}
\end{table}

We also show in  Table~\ref{tab:chi2} the contribution of each experiment to the best-fit $\chi^2_\mathrm{eff}$, which can be compared to the $\Lambda$CDM case. We find that most of the improvement is driven by the Planck SZ cluster data, which can be fitted perfectly by the IDS model, instead of being discrepant at the 3.9$\sigma$ level. About 15 units of $\Delta \chi^2$ come from there. Next, the two limits of the IDS model with small $\dN$ provide a slightly better fit to Planck high-$\ell$ TT+TE+EE data, by about $\Delta \chi^2 \simeq -5$. Improvements in other data sets are not significant.

Note that our best-fit WI, DP and general IDS models do not improve the fit to direct measurements of $H_0$ over that of the $\Lambda$CDM model. This discrepancy contributes a $\chi^2$ ranging from 8.4 to 9.4, i.e. 2.9 to 3.1$\sigma$ away from the measured central value. On the contrary, the WI model with a linear prior $\dN \geq 0.07$ allows for a significantly better fit to $H_0$ with a $\chi^2=4$ for this data point (2$\sigma$ away from the central value).

These results are statistically consistent since the minimum $\chi^2$ goes down when the model is more general. The WI model with $\dN \geq 0.07$ is a subcase of the WI model with $-5 \leq \log_{10} \dN \leq 0$, which is itself a subcase of the general IDS model; and the same is true with the DP models. The minimum $\chi^2$'s are ordered accordingly. This does not imply that the $\chi^2$ of each experiment at the best-fit point should respect this order. For instance, among our best-fit models, the one with the smallest $\chi^2$ for the $H_0$ data point is the WI model restricted to $\dN \geq 0.07$. This is not inconsistent: it results from the pulls of different experiments which remain in slight tension with each other along different directions in parameter space. We can anticipate from these results that it is difficult to provide simultaneously a better fit to Planck SZ and to $H_0$ data. When very small values of $\dN$ are allowed, the Planck SZ data push towards small $\sigma_8$ values at the expense of a nearly constant $H_0$; while with $\dN \geq 0.07$ the data favors a compromise between the $\sigma_8$ and $H_0$ values.

Finally, in Table~\ref{tab:chi2_gen} we show the minimum value of $\chi^2_\mathrm{eff} =-2\ln {\cal L}$ for both \LCs and the general IDS model with $\log$ priors on the three parameters $\boldsymbol{\theta}_\mathrm{IDS}$, with the same experiments. In this case, the posterior parameter probability distributions are strongly non-gaussian, and we switched the parameter extraction method in MontePython to MultiNest mode. For faster convergence, we reduced the number of nuisance parameters and used the ``Planck lite'' version of the high-$\ell$ Planck2015 likelihood.

After these preliminary comments on the best-fit models, we must look at the confidence limits on each parameter to better understand what the data is telling us.

\subsubsection{The parameters}

The mean values and confidence limits for each parameters are given in Table~\ref{tab:CLs_lin} for runs with a linear prior $\dN \geq 0.07$, and in
Table~\ref{tab:CLs_log} for runs with a log prior $-5 \leq \log_{10} \dN \leq 0$.

\subsection{Results with a linear prior $\dN \geq 0.07$}

In this case, the data prefers a non-zero scattering rate $\Gamma_0$ in the WI model at the 3.4$\sigma$ level (respectively, a non-zero fraction of interacting DM $f$ in the DP model at the 3$\sigma$ level). These levels of significance are consistent with the $\Delta \chi^2_\mathrm{eff}$ discussed in the previous section. The WI models have a mean value of $\Gamma_0 \simeq 1.1\times10^{-7}$Mpc$^{-1} \simeq 1.1\times 10^{-21}$s$^{-1}$, compatible with the 2015 results of~\cite{Lesgourgues:2015wza} at the 1.5$\sigma$ level, and the DP models have a mean IDM fraction of 1.4\%. We recall that by construction, $\Gamma_0$ is not constrained by the data in the DP model, since this model is defined as the limit in which the interaction is very efficient ($\Gamma_0 \gg H_0$) and its precise value does not matter.

For both the WI and DP models, $\dN$ is constrained from below by the theoretical prior $\dN \geq 0.07$, and from above by the data at the level of $\dN \sim 0.67$ for WI (resp. 0.51 for DP) at the 95\%CL level. In the 2015 results of~\cite{Lesgourgues:2015wza}, high values of $H_0$ could be reached for models with a significant DR density. It appears that the more recent CMB and LSS data used in this analysis better constrains the DR density, by about 50\%, and reduces the possibility to reach high Hubble parameter values.  Indeed, the confidence intervals obtained for $H_0$ when fitting our whole data set (including the Hubble data point from \cite{Riess:2016jrr}) are only compatible with that measurement at the 1.5$\sigma$ level for WI, or 2.0$\sigma$ level for DP, to be compared with the 2.6$\sigma$ level for $\Lambda$CDM.

In summary, the IDS models can only render the Hubble tension more moderate, but the Planck SZ data can be much better fitted, and drives some $\sim 3\sigma$ evidence for the presence of interacting dark matter, together with a DR density close to the lower prior edge $\dN \sim 0.07$.

\begin{table}
	\begin{tabular}{|| c || c | c | c ||}
		\hline
		\multicolumn{4}{|c|}{Parameter mean values and 68\%CL confidence interval (or 95\%CL upper limit), lin. priors} \\
		\hline\hline
		Parameters & \LC & WI limit & DP limit \\ [0.5ex] 
		\hline\hline
		$100\ob$ & $2.245_{-0.014}^{+0.013}$ & $2.249_{-0.019}^{+0.018}$ & $2.242_{-0.019}^{+0.017}$ \\
		\hline
		$n_\so$ & $0.9656_{-0.0037}^{+0.0038}$ & $0.9708_{-0.0041}^{+0.0044}$ & $0.9701_{-0.0042}^{+0.0038}$ \\
		\hline
		$\tau_\mathrm{reio}$ & $0.04887_{-0.008}^{+0.008}$ & $0.05915_{-0.0078}^{+0.0082}$ & $0.06118_{-0.0086}^{+0.0093}$ \\
		\hline
		$\h0$ & $68.67_{-0.46}^{+0.41}$ & $70.01_{-1.2}^{+1.1}$ (95\% CL: $72.21$) & $69.13_{-1.3}^{+0.76}$ (95\% CL: $71.32$) \\
		\hline
		$\ln10^{10}A_{s }$ & $3.023_{-0.015}^{+0.015}$ & $3.05_{-0.017}^{+0.017}$ & $3.056_{-0.019}^{+0.022}$ \\
		\hline
		$\odmtot$ & $0.1168_{-0.00089}^{+0.001}$ & $0.126_{-0.0039}^{+0.0032}$ & $0.1235_{-0.0033}^{+0.0017}$ \\
		\hline\hline
		$\dN$ & 0 & $0.369_{-0.19}^{+0.17}$ (95\% CL: $\leq 0.6657$) & $\leq 0.5064$ (95\% CL) \\
		\hline
		$10^7 \g0$ & 0 & $1.097_{-0.32}^{+0.32}$ & $\g0 \gg \h0$ \\
		\hline
		$f$ & 0 & 1 & $0.01387_{-0.0046}^{+0.0052}$ \\
		\hline\hline\hline
		$100\theta_\so$ & $1.042_{-0.0003}^{+0.00028}$ & $1.043_{-0.00037}^{+0.00035}$ & $1.043_{-0.00038}^{+0.00036}$ \\
		\hline
		$\s8$ & $0.7933_{-0.0054}^{+0.0052}$ & $0.7721_{-0.01}^{+0.01}$ & $0.7734_{-0.012}^{+0.011}$ \\
		\hline
		$\Omega_\ma$ & $0.2968_{-0.0053}^{+0.0057}$ & $0.3043_{-0.0053}^{+0.0067}$ & $0.3067_{-0.007}^{+0.0074}$ \\ [1ex]
		\hline\hline
	\end{tabular}
	\caption{Parameter mean values and 68\%CL confidence interval (or 95\%CL upper limit), in the WI and DP cases, with linear priors on all parameters.\label{tab:CLs_lin}}
\end{table}

\subsection{Results with a logarithmic prior $-5 \leq \log_{10} \dN \leq 0$}

In this case, the data prefers a non-zero scattering rate $\Gamma_0$ in the WI model at the 2.9$\sigma$ level (respectively, a non-zero fraction of interacting DM $f$ in the DP model at the 3.0$\sigma$ level). At the same time, it favors small values of the DR density that were previously excluded by the prior.

The WI models have a mean value of $\Gamma_0 \simeq 2.3\times10^{-7}$Mpc$^{-1} \simeq 2.3\times 10^{-21}$s$^{-1}$, and of $\dN\simeq 0.0049$. The DP models have a mean IDM fraction of 4.8\%, and of $\dN\simeq 0.0015$. Note that the data cannot be sensitive to the direct effect of such low DR densities. However, to get the right amount of DR drag on DM and the right shape for the MPS and CMB spectrum, at least some DR is required.

The models are driven to this new region in parameter space mainly by the Planck SZ data, which can be extremely well fitted in that case. Interestingly, the CMB temperature and polarization data can also be slightly better fitted, see Table~\ref{tab:chi2}. This is done at the expense of fitting a high $H_0$: with such low values of $\dN$, the confidence intervals on $H_0$ do not change significantly compared to the $\Lambda$CDM case, and the level of tension is the same.

In summary, the models most favored in this analysis have a tiny DR density, and at the same time larger values of $\Gamma_0$ or $f$. The combined effects from the IDS gives a very significant improvement in the goodness-of-fit ($\Delta \chi^2 \sim 20$) driven mainly by Planck SZ data and secondarily by Planck CMB high-$\ell$ data, but without easing the tension with direct Hubble measurements. Posteriors and likelihood contours for the cosmological parameters of the WI, DP, and general IDS models are shown in figures \ref{fig_7}, \ref{fig_8} and \ref{fig_9}.

\begin{table}
	\begin{tabular}{|| c || c | c | c | c ||}
		\hline
		\multicolumn{5}{|c|}{Parameter mean values and 68\%CL confidence interval, $\log \dN$ prior} \\
		\hline\hline
		\multirow{2}{*}{Parameters} & \multirow{2}{*}{\LC} & WI limit & DP limit & General IDS \\ [0.5ex] 
		 & & \tiny{$\log \dN$ Prior} & \tiny{$\log \dN$ Prior} & \tiny{$\log\dN$, $\log\Gamma_0$, $\log f$ $\log$ Priors} \\
		\hline\hline
		$100\ob$ & $2.245_{-0.014}^{+0.013}$ & $2.228_{-0.012}^{+0.012}$ & $2.231_{-0.014}^{+0.014}$ & $2.235_{-0.013}^{+0.013}$ \\
		\hline
		$n_\so$ & $0.9656_{-0.0037}^{+0.0038}$ & $0.9625_{-0.0033}^{+0.0039}$ & $0.9628_{-0.0035}^{+0.0035}$ & $0.9670_{-0.0037}^{+0.0035}$ \\
		\hline
		$\tau_\mathrm{reio}$ & $0.04887_{-0.008}^{+0.008}$ & $0.05815_{-0.0077}^{+0.0078}$ & $0.05827_{-0.0082}^{+0.0082}$ & $0.05835_{-0.0077}^{+0.0079}$ \\
		\hline
		$\h0$ & $68.67_{-0.46}^{+0.41}$ & $67.84_{-0.3}^{+0.42}$ & $67.98_{-0.38}^{+0.35}$ & $68.06_{-0.42}^{+0.39}$ \\
		\hline
		$\ln10^{10}A_{s }$ & $3.023_{-0.015}^{+0.015}$ & $3.047_{-0.016}^{+0.015}$ & $3.047_{-0.016}^{+0.016}$ & $3.049_{-0.016}^{+0.017}$ \\
		\hline
		$\odmtot$ & $0.1168_{-0.00089}^{+0.001}$ & $0.119_{-0.001}^{+0.00065}$ & $0.1185_{-0.00081}^{+0.00084}$ & $0.1184_{-0.00089}^{+0.00087}$ \\
		\hline\hline
		$\log_{10} \dN$ & \textendash & $-2.309_{-0.25}^{+0.25}$ & $-2.814_{-0.19}^{+0.2}$ & $-2.67_{-0.32}^{+0.27}$ \\
		\hline
		$10^7 \g0$ & 0 & $2.272_{-0.65}^{+0.61}$ & $\g0 \gg \h0$ & $\log_{10} \g0 \Mpc > -7.1$ (95\% CL) \\
		\hline
		$f$ & 0 & 1 & $0.04785_{-0.017}^{+0.016}$ & $\log_{10} f > -1.7$ (95\% CL) \\
		\hline\hline\hline
		$100\theta_\so$ & $1.042_{-0.0003}^{+0.00028}$ & $1.042_{-0.00026}^{+0.00027}$ & $1.042_{-0.00027}^{+0.00029}$ & $1.0418_{-0.00041}^{+0.00042}$ \\
		\hline
		$\s8$ & $0.7933_{-0.0054}^{+0.0052}$ & $0.7565_{-0.0092}^{+0.01}$ & $0.7588_{-0.011}^{+0.0099}$ & $0.762_{-0.012}^{+0.012}$ \\
		\hline
		$\Omega_\ma$ & $0.2968_{-0.0053}^{+0.0057}$ & $0.3083_{-0.0057}^{+0.0039}$ & $0.3062_{-0.0048}^{+0.0048}$ & $0.305_{-0.0052}^{+0.0051}$ \\ [1ex]
		\hline\hline
	\end{tabular}
	\caption{Parameter mean values and 68\%CL confidence interval (or 95\%CL upper limit), in the WI, DP and general IDS cases, with a logarithmic prior on $dN$. \label{tab:CLs_log}}
\end{table}

{\centering
	\begin{figure}
		\includegraphics[width=0.9\textwidth]{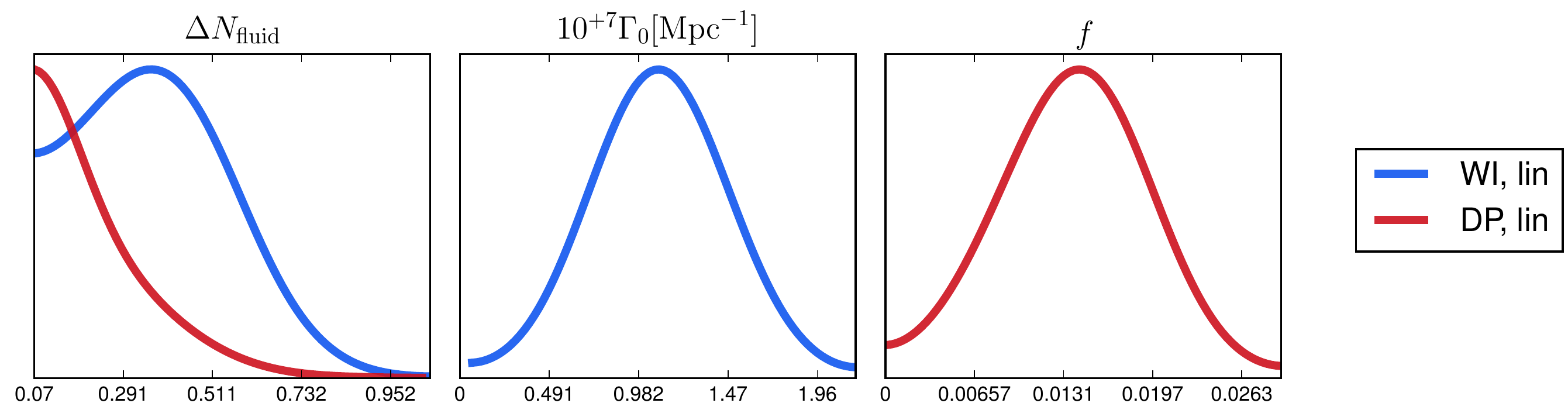}
		\includegraphics[width=0.9\textwidth]{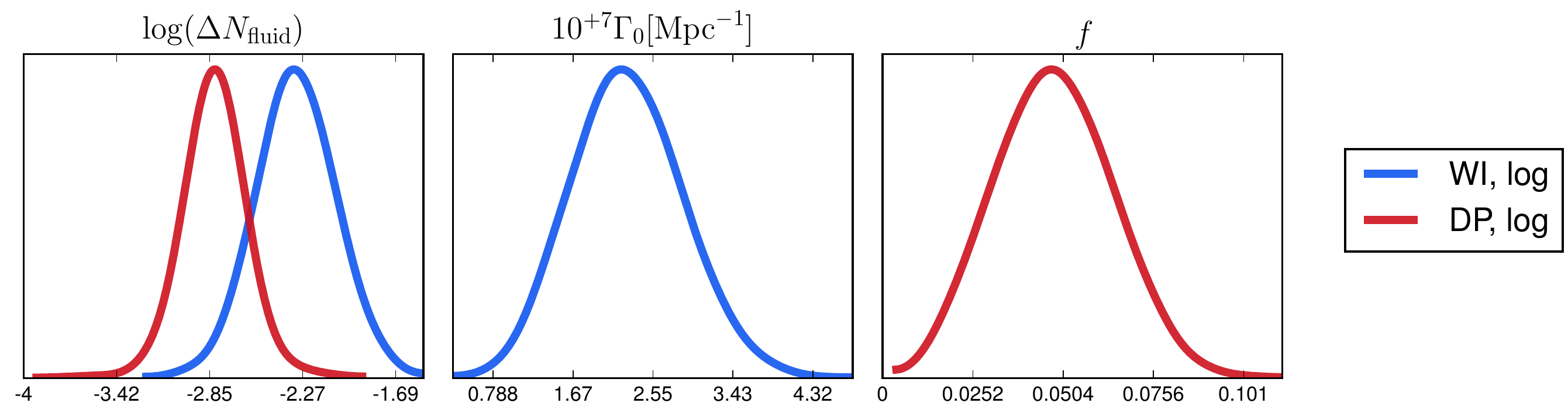}
		\caption{Posteriors of the IDS parameters for $\LC$, WI and DP with (top) a $\dN \geq 0.07$ prior and (bottom) a $-5 \leq \log_{10} \dN \leq 0$ prior.}\label{fig_7}
	\end{figure}
}

{\centering
	\begin{figure}
		\includegraphics[width=0.45\textwidth]{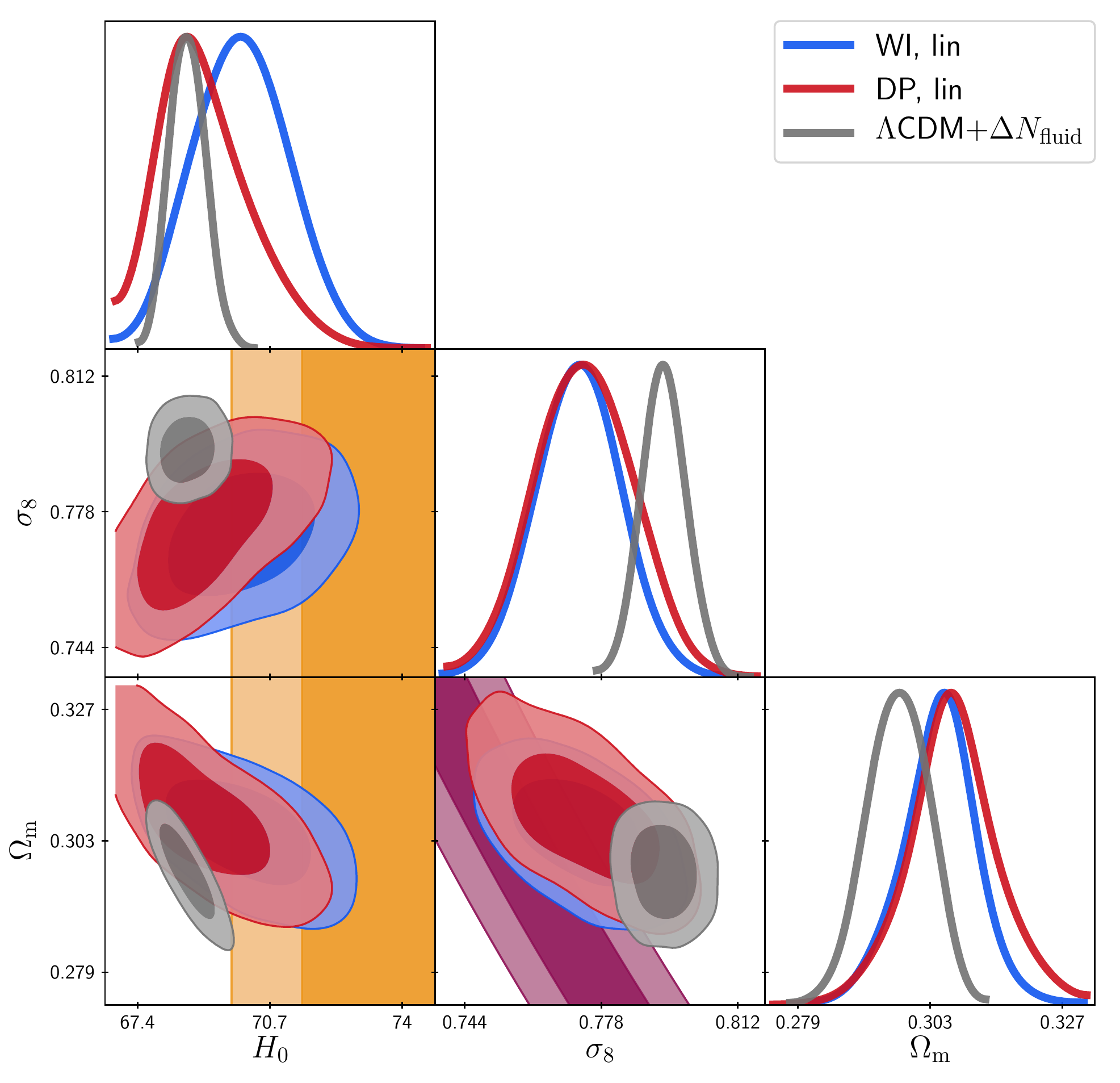}
		\includegraphics[width=0.45\textwidth]{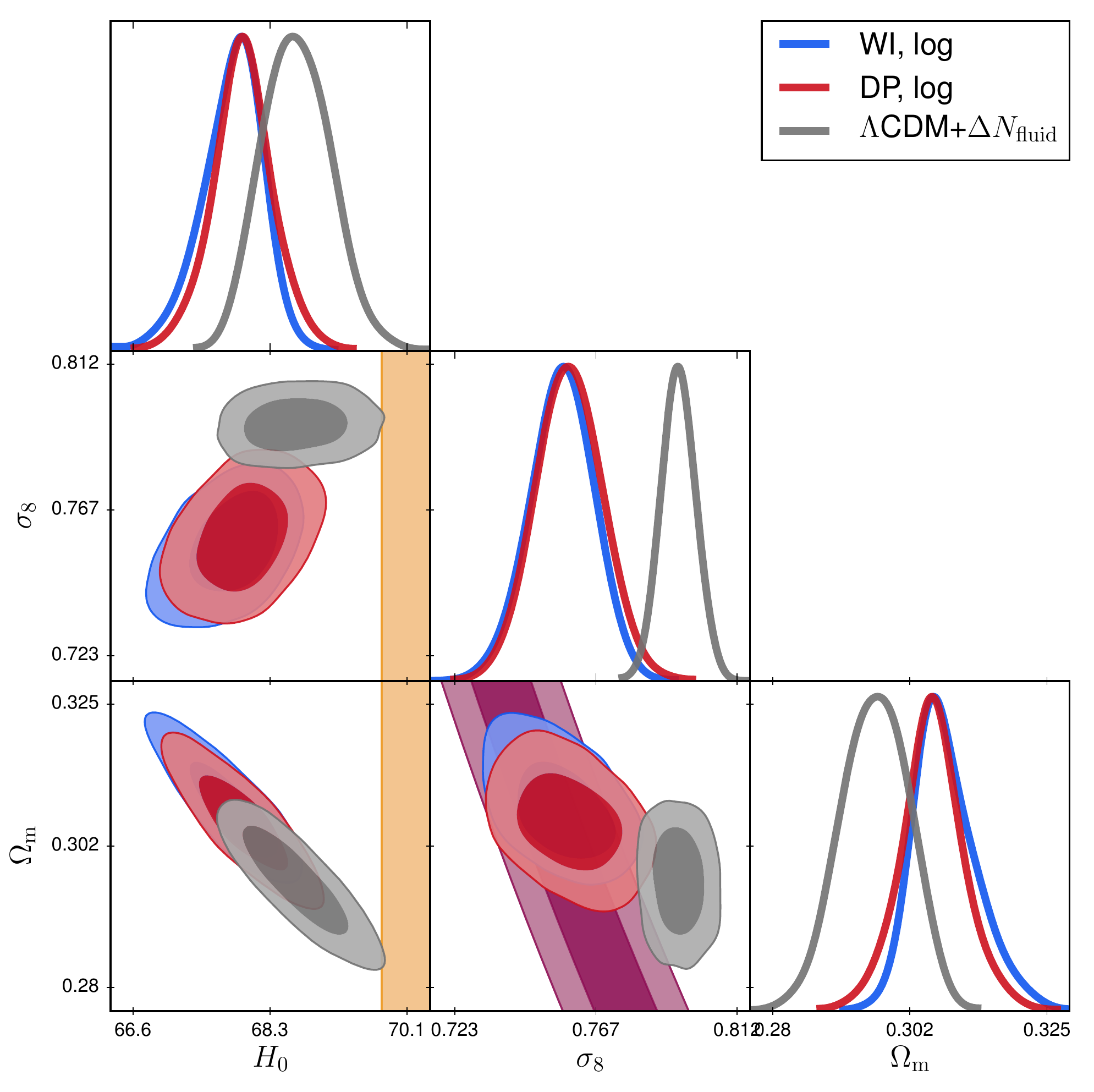}
		\caption{Posteriors and likelihood contours for $\LC$, WI and DP with (left) a $\dN \geq 0.07$ prior and (right) a $-5 \leq \log_{10} \dN \leq 0$ prior.
		On the right plot, the WI and DP models are almost indistinguishable. The orange contours show the $H_0$ measurement by Riess et al.~\cite{Riess:2016jrr}, 
		and the purple ones the $\sigma_8 (\Omega_\ma / 0.27)^{0.30}$ constraint from Planck SZ cluster counts \cite{Ade:2013lmv}.}\label{fig_8}
	\end{figure}
}

{\centering
	\begin{figure}
		\includegraphics[width=0.45\textwidth]{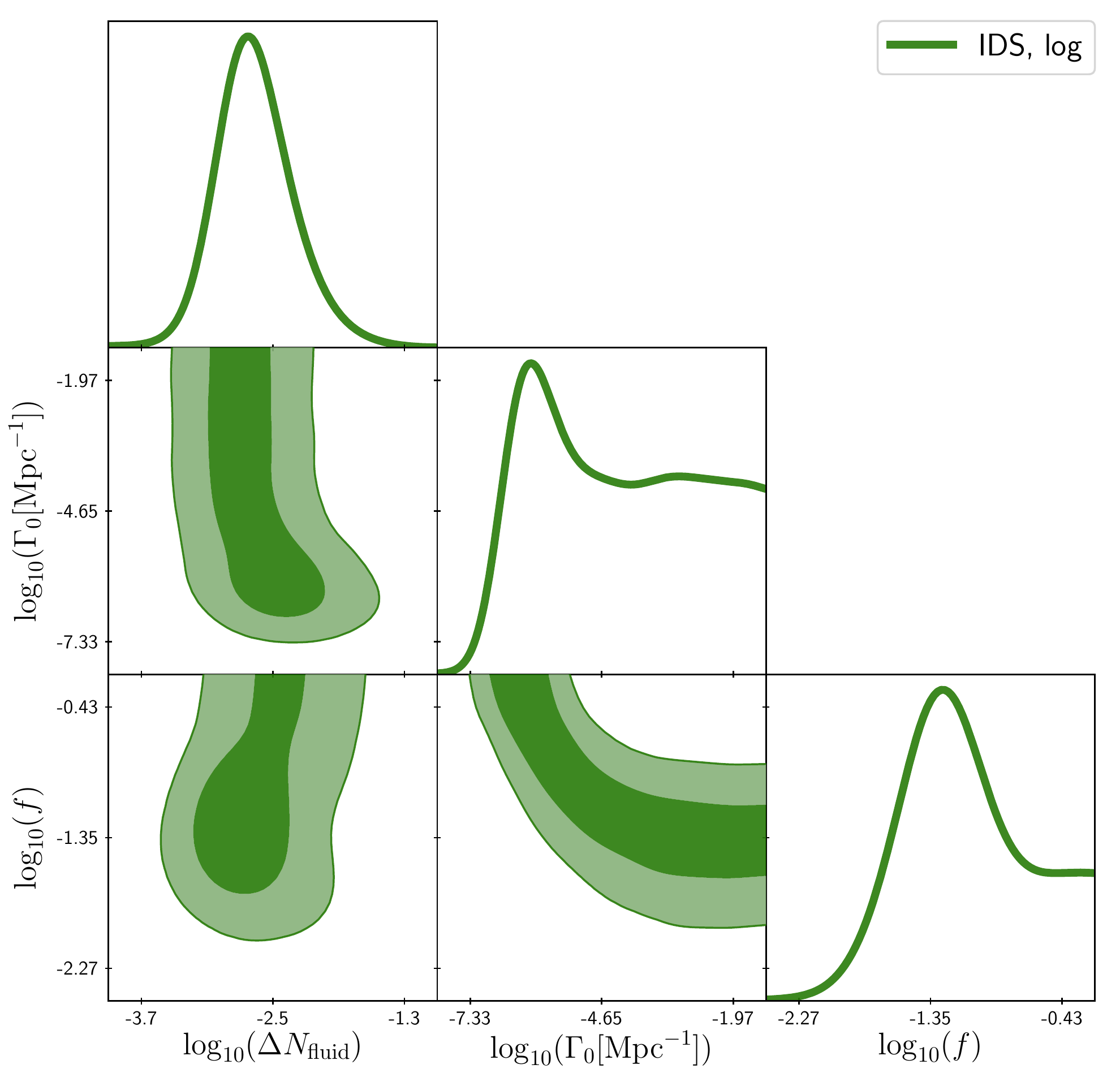}
		\includegraphics[width=0.45\textwidth]{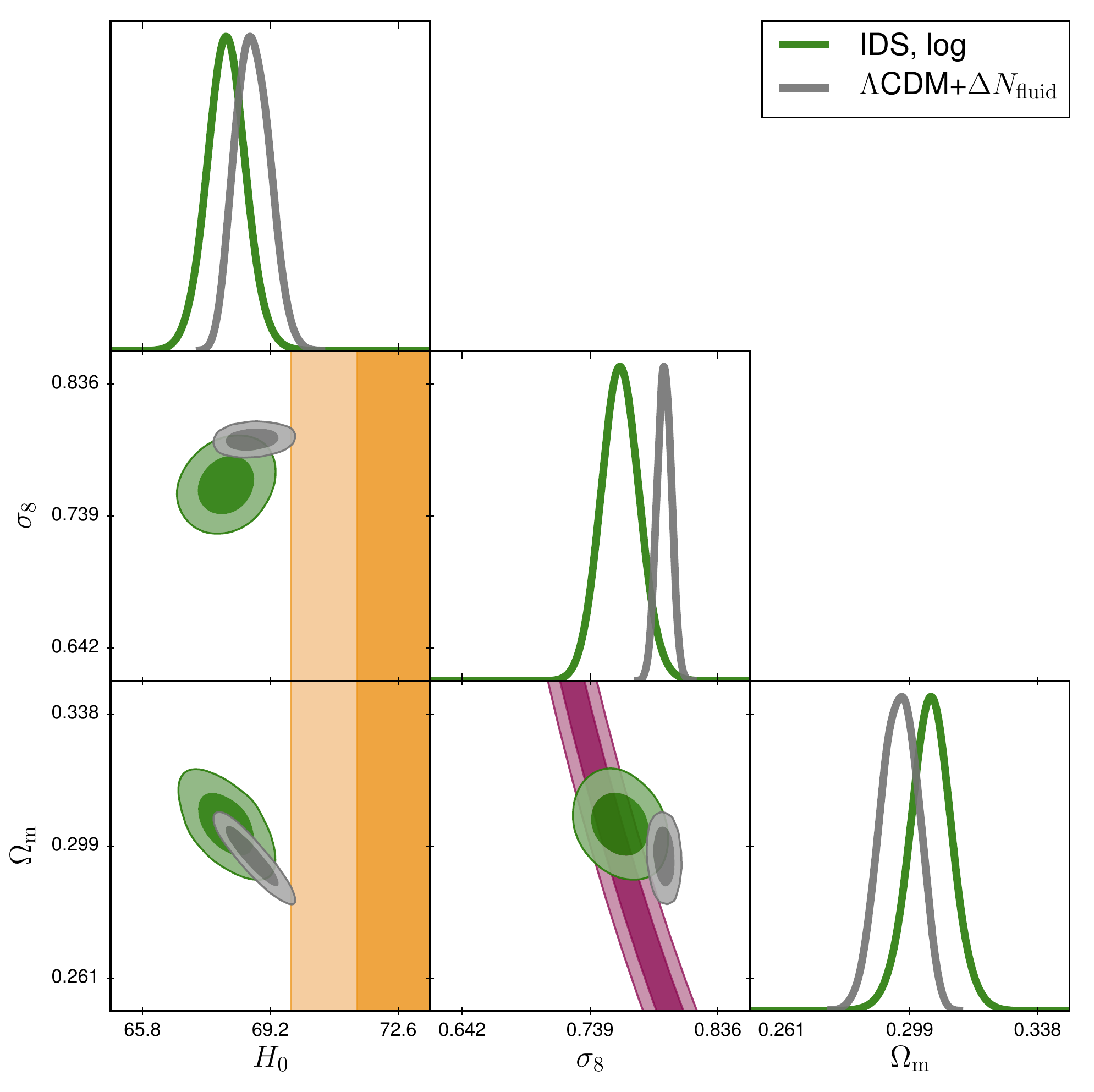}
		\caption{Posteriors and likelihood contours for $\boldsymbol{\theta}_\mathrm{IDS}$ in the general IDS model with $\log$ priors on $\boldsymbol{\theta}_\mathrm{IDS}$ (left), and the posteriors and likelihood contours for $\{ \h0, \s8, \Omega_\ma \}$ for this model and $\LC$ (right). Note how on the left plot the WI and DP limits can be observed in the $\log_{10} f$ vs. $\log_{10} \g0$ contour plot, as the ends of a "canyon" region of values that represents a family of models that provide a good fit to the data. On the right plot, the orange contours show the $H_0$ measurement by Riess et al.~\cite{Riess:2016jrr}, and the purple ones the $\sigma_8 (\Omega_\ma / 0.27)^{0.30}$ constraint from Planck SZ cluster counts \cite{Ade:2013lmv}.}\label{fig_9}
	\end{figure}
}

\section{Conclusions and Comments}
\label{sec:conc}

In this paper, we have studied a class of cosmological models that allow a fraction of the dark matter to interact with a locally thermal dark radiation fluid, thus generalizing the works of \cite{Buen-Abad:2015ova,Lesgourgues:2015wza,Chacko:2016kgg}. These interacting Dark Sector (IDS) models have three parameters in addition to the usual six of \LC: the amount $\rho_\dr$ of dark radiation, the fraction $f$ of the total DM that is interacting with this radiation, and the interaction rate $\Gamma = \g0 a^{-2}$. Previous work has focused on either one of two well-motivated limits in which only two parameters are relevant. The weakly interacting (WI) limit in which all the DM is interacting ($f=1$, $\Gamma \ll H$; \cite{Buen-Abad:2015ova,Lesgourgues:2015wza}) or the Dark Plasma (DP) limit ($f < 1$, $\g0 \gg \h0$; \cite{Schmaltz,Chacko:2016kgg}) in which the coupling is strong but in which only a fraction of the dark matter interacts with the dark radiation.  

In either case, the dark matter - dark radiation interactions reduce the rate of growth of matter perturbations and thus result in a suppression of the Matter Power Spectrum (MPS). The reduction of the rate of growth in the two limiting cases has a different time and wave vector dependence leading to differing predicted shapes of the MPS as a function of $k$ and $z$. Thus general IDS model which includes both of these limits gives rise to a family of MPS all with the same predicted value of $\sigma_8$ but with smoothly varying shapes (in $z$, $k$) as a function of the model parameters $f$ and $\Gamma$.   

In order to determine whether the IDS models are preferred over \LC, and whether existing data already prefer a particular shape for the suppression of the MPS, we fitted the general IDS model (and the two limits) to cosmological data from the CMB and BAO, as well as from LSS experiments and local measurements of the expansion rate of the Universe. The LSS experiments favor a smaller MPS than the one predicted by \LCs from CMB and BAO data (the \s8 problem), while the measurements of the expansion rate of the Universe favor a larger Hubble parameter (the \h0 problem).

We found that the IDS models significantly improved the global fit to the full data set ($3-4\sigma$), solving the \s8 problem and in some cases also relieving the tension in the \h0 problem. We defer the inclusion of recent KiDS and DES weak lensing likelihoods in our analysis to future work. We expect that this will even strengthen our conclusions.
Indeed, when the KiDS-450 + 2dFLenS weak lensing and redshift-space galaxy clustering data is reduced  to a gaussian constraint on $\sigma_8 (\Omega_{\rm m} / 0.30)^{0.5} = 0.742 \pm 0.035$~\cite{Joudaki:2017zdt}, we find that the five IDS best-fit models obtained in this analysis have a $\chi^2$ of 0.25 to 1.9 with respect to this data point, thus solving the 2.6$\sigma$ tension with Planck  $\Lambda$CDM claimed in~\cite{Joudaki:2017zdt}. Reducing the DES-Year-1 shear, galaxy and cross-correlation data to a single measurement of $\sigma_8 (\Omega_{\rm m} / 0.30)^{0.5} = 0.783^{+0.021}_{-0.025}$~\cite{Abbott:2017wau}, the same best-fit models have $\chi^2$'s from 0 to 1.0.

Unfortunately, current LSS experiments are not yet sensitive enough to significantly favor either of the shapes predicted by the WI and DP limits. Data sets based on observation of visible matter as tracers for the matter power spectrum are currently limited by systematics associated with galaxy bias (i.e. modeling of how well visible matter traces DM) while the in-principle less biased weak lensing data is not yet precise enough to distinguish the different shapes.   
Future improvements in these experiments, as well as the onset of new probes like the one making use of the $21$-cm hydrogen line with the possibility to probe the MPS at different redshifts will dramatically change this situation. Such measurements promise to turn the matter power spectrum into a precision cosmological tool that can shed light on any shady business that might be going on in the dark sector and distinguish between different models with DM interactions.

Measurements of the flux power spectrum of Lyman-$\alpha$ forests in quasar spectra are also potentially very sensitive to the IDS effects.
Strictly speaking, one is not allowed to use current Lyman-$\alpha$ likelihood for the IDS model, because the latter features a dark matter scale-dependent growth rate different from that of the $\Lambda$CDM (or $\Lambda$CDM + massive neutrino) models. 
Hence it would be necessary to run dedicated hydrodynamical simulations for IDS models, and include the results in the analysis pipeline.
Nonetheless, one can ignore these complications to at least get a rough idea of the impact of Lyman-$\alpha$ data on our model.
This has been tried by Krall et al.~\cite{Krall:2017xcw}, using some pioneering SDSS Lyman-$\alpha$ data from 2004~\cite{McDonald:2004eu,McDonald:2004xn}, which consist of a joint distribution of probability for the amplitude and slope of the linear power spectrum at scales to which the flux power spectrum is maximally sensitive.
The analysis of~\cite{Krall:2017xcw} shows that the inclusion of 2004 SDSS Lyman-$\alpha$ data reduces the goodness of fit of the Weakly Interacting model.
We agree with this conclusion, which is consistent with the fact that the 2004 SDSS Lyman-$\alpha$ data analysis returned a  rather large value of $\sigma_8$, 
with a central value of 0.85~\cite{McDonald:2004xn}, compatible with the Planck $\Lambda$CDM best fit model.
The BOSS  flux power spectrum measurement of \cite{Palanque-Delabrouille:2013gaa} is based on more recent data and on a different treatment of systematics and nuisance parameters.
It would be interesting to perform hydrodynamical simulations and use this data for the IDS model.
We anticipate that the 2016 BOSS Lyman-$\alpha$ data might bring further support for the IDS, because it has a lower $\sigma_8$ than the 2004 SDSS Lyman-$\alpha$ data, and, interestingly, it prefers a lower value of the spectral index $n_s$ than the Planck $\Lambda$CDM best fit  model (see e.g. Figure~8 in \cite{Yeche:2017upn}). This constraint on $n_s$ applies essentially to small scales at which the flux power spectrum is measured. The IDS model (and in particular its WI limit) can be thought of as a way to lower the effective $n_s$ in the small-scale power spectrum, while keeping the concordance value at large scales tested by CMB data. Hence it could probably explain the low $n_s$ of the BOSS Lyman-$\alpha$ data.

In this work, we have ignored the effects of self-interactions of the IDM. Quantum field theory requires that any particle physics model which contains interactions between IDM and DR necessarily also has scattering between IDM particles themselves. In the weakly coupled limit these IDM-IDM interactions are too small to be relevant. However in the DP limit the coupling can be large and the IDM component of the DM can become strongly self-interacting. Such IDM self-interactions might responsible for the small scale discrepancies observed in dark matter halos or they could give rise to interesting DM halo substructures such as a DM disk if they allow dissipation through IDM scattering with associated DR emission (see for example \cite{Fan:2013yva}).

\section*{Acknowledgments}
\label{sec:ack}

We thank Andy Cohen, Sungwoo Hong, Gustavo Marques-Tavares, Luke Pritchett, Yuhsin Tsai, Evan Weinberg and Yiming Zhong for helpful comments and discussions. The work of MBA and MS is supported by DOE grant DE-SC0015845. MS would like to thank the Institute for Advanced Studies at HKUST and the Aspen Center for Physics (which is supported by NSF grant PHY-1607761) for hospitality during work on this project. This work used the facilities offered by the RWTH High Performance Computing Cluster from Aachen University, as well as the computing resources offered by the Boston University Shared Computing Cluster (SCC) in the MGHPCC.

\textit{Note added.\textendash} After the publication of v1 of our paper, \cite{Raveri:2017jto} appeared, in which the Dark Plasma limit of our model is fit to different set of LSS data. The conclusions from their work are in qualitative agreement with ours.

\appendix

\section{Perturbation equations from Boltzmann equation with DM-DR interactions}
\label{appA}

In this Appendix we elaborate on some aspects of the derivation of the perturbation equations governing the IDM and DR fluids (\Eqs{thdr_1}{R_def}) which we first presented in \cite{Buen-Abad:2015ova}. Subsequent to \cite{Buen-Abad:2015ova}, \cite{Cyr-Racine:2015ihg} derived perturbation equations governing IDM and DR in the ETHOS formalism, allowing for general momentum dependence of the IDM-DR scattering matrix element. Our model is a special case of this formalism and thus our perturbation equations should be obtainable from the formulas in \cite{Cyr-Racine:2015ihg}. However, there is a subtlety which led to a disagreement between our results and those published in \cite{Cyr-Racine:2015ihg}. Since then, the authors of  \cite{Cyr-Racine:2015ihg} have replaced their paper (new version v4), and their results now agree with ours. This Appendix contains a discussion of the subtle points in the derivation which we hope will be useful to the interested reader. Throughout this Appendix when we quote \cite{Cyr-Racine:2015ihg} we refer to version v3 of the paper. 

Specifically, the discrepancy between our results and those of \cite{Cyr-Racine:2015ihg} was a 3/2 instead of our 3/4 in the factor $R \equiv {3 \rho_\idm}/{4 \rho_\dr}$ that relates the interaction terms in the IDM and DR equations (see \Eq{R_def}). Here, we re-derive the perturbation equations starting from the Boltzmann equations. We closely follow the derivation in \cite{Cyr-Racine:2015ihg} (for pedagogical introductions see \cite{Ma:1995ey} and \cite{Dodelson:2003ft}). We find that the discrepancy arises from an unusual definition of the DR density and velocity perturbations in \cite{Cyr-Racine:2015ihg} which obscures the conservation of energy and momentum. Version v4 of \cite{Cyr-Racine:2015ihg} reverts to the conventional definition of the DR density and velocity perturbations and obtains results which agree with ours.

\subsection{The Boltzmann equations}
The objects of interest in the Boltzmann formalism are the phase space distribution functions of the IDM and the DR which we denote by $f_\idm$ and $f_\dr$. The distribution functions $f(\mathbf{x}, \mathbf{p}, \eta)$ describe the probability of finding a particle with 3-momentum $\mathbf{p}$ at the location $\mathbf{x}$ as a function of conformal time $\eta$. It is convenient to write the distribution functions as
\begin{equation}\label{appA1}
f(\mathbf{p}, \mathbf{x}, \eta) = \f0(p,\eta) \bl( 1 + \Psi ( \mathbf{p},\mathbf{x}, \eta) \br) \ ,
\end{equation}
where the zeroth-order distribution function $\f0(p, \eta)$ is independent of $\mathbf{x}$, reflecting that the Universe is approximately homogeneous and isotropic. $\Psi$ (denoted by $\Theta$ in \cite{Cyr-Racine:2015ihg}) describes the small perturbations about the homogeneous solution which we wish to derive an equation for. For linear perturbations it is more convenient to work in Fourier space where $\Psi(\mathbf{p}, \mathbf{k}, \eta)$ is a function of the Fourier wavenumber $\mathbf{k}$. Finally, focusing only on scalar perturbations, it can be shown that $f$ only depends on the angle between $\mathbf{p}$ and $\mathbf{k}$ and the magnitudes $p$, $k$ so that  $\Psi = \Psi (p, \mu \equiv \hat{p} \cdot \hat{k}, k, \eta)$ where $p$ and $k$ are the magnitudes of $\mathbf{p}$ and $\mathbf{k}$ and $\mu$ is the cosine of the angle between them.

The Boltzmann equation determining the evolution of the IDM and DR distribution functions can formally be written as
\begin{equation}\label{appA2}
\hat{L} [f_i] = \hat{C}_i[f_\idm, f_\dr] \ ,
\end{equation}
where $\hat{L}$ is the Liouville operator, and $\hat{C}$ is a collision operator which describes the collisions between the IDM and the DR particles. This equation can be made more explicit for a perturbed FLRW metric (see for example \cite{Ma:1995ey, Dodelson:2003ft} or \cite{Cyr-Racine:2015ihg}). We quote its form in Newtonian gauge ($\phi$ and $\psi$ are the gravitational potentials)
\begin{eqnarray}
\dot{f}_\idm + i k \frac{p}{E} \hat{p} \cdot \hat{k} \ f_\idm +  p \frac{\partial f_\idm}{\partial p} \bl( -\mH + \dot{\phi} - i k \frac{E}{p} \hat{p} \cdot \hat{k} \ \psi \br) & = & \frac{a}{E} (1+\psi) \Big[ C_{\idm-\dr}(\mathbf{p}) \nn\\
& & + C_{\idm-\idm}(\mathbf{p}) \Big] \ , \label{appA3} \\
\dot{f}_\dr + ik \hat{q} \cdot \hat{k} \ f_\dr + q \frac{\partial f_\dr}{\partial q} \bl( \dot{\phi} - i k \hat{q} \cdot \hat{k} \ \psi \br) & = & \frac{a^2}{q} (1+\psi) \Big[ C_{\dr-\idm}(\mathbf{q}/a) \nn\\
& & + C_{\dr-\dr}(\mathbf{q}/a) \Big] \ ; \label{appA4}
\end{eqnarray}
where $\mathbf{q} \equiv a \mathbf{p}$ is the comoving momentum of the DR; and $C_{ij}(\mathbf{p}_i)$ denotes the $ij \rightarrow ij$ collision term for particles of species $i$ with momentum $\mathbf{p}_i$.
In principle, these differential equations must integrated to obtain the distribution functions $f_\idm$ and $f_\dr$ which contain a full description of
the IDM and DR fluids. 

However, in practice one is usually only interested in certain macroscopic quantities for each fluid which correspond to moments of the distribution functions and which are much easier to obtain. The most important macroscopic quantities needed for the cosmological linear perturbation equations for each fluid $i$ are the density $\delta$ and velocity $\theta$ perturbations defined as \cite{Ma:1995ey}:
\begin{eqnarray}
\overline{\rho}_i(\eta) (1 + \delta_i (\mathbf{k}, \eta)) & \equiv & g_i \int \frac{\text d^3 \mathbf{p}}{(2 \pi)^3} \,\, \f0_i ( 1 + \Psi_i ) \ E \ , \label{appA5} \\
\overline{P}_i(\eta) + \delta P_i(\mathbf{k}, \eta) & \equiv & g_i \int \frac{\text d^3 \mathbf{p}}{(2 \pi)^3} \,\, \f0_i (1 + \Psi_i) \ \frac{p^2}{3E} \ , \label{appA6} \\
(\overline{\rho}_i+\overline{P}_i) \theta_i(\mathbf{k}, \eta) & \equiv & g_i \int \frac{\text d^3 \mathbf{p}}{(2 \pi)^3} \,\, \f0_i \Psi_i \ (i \mathbf{k} \cdot \mathbf{p}) \ ; \label{appA7}
\end{eqnarray}
where $g_i$ counts the number of internal degrees of freedom of the IDM or the DR, and $E$, $p$ are the energy and momentum of the particles. From now on we assume stable and highly non-relativistic DM so that $E_\idm \approx M_\idm \equiv M$ and massless DR, $E_\dr = p_\dr$. Expanding the left- and right-hand sides of these equations to zeroth and first order defines the average energy and momentum densities $\overline \rho, \overline P$, as well as the perturbations $\delta , \theta$. 

Taking the appropriate moments (i.e. momentum integrals) of \Eqs{appA3}{appA4} yields the evolution equations obeyed by the macroscopic quantities described in \Eqsto{appA5}{appA7}. A corollary of energy-momentum conservation in particle scattering is that self-scattering of particles within one fluid cannot change the energy or momentum densities of that fluid (even though it can change the distribution function). This implies that 
the $E$ and $\mathbf{k} \cdot \mathbf{p}$ moments of the self-interaction collision terms $C_{ii}$ on the r.h.s. of \Eqs{appA7}{appA8} vanish
\begin{eqnarray}
\int \text d^3 \mathbf{p} \,\, E \bl( \frac{1}{E} C_{ii} (\mathbf{p}) \br) & = & 0 \ , \label{appA8} \\
\int \text d^3 \mathbf{p} \,\, (\mathbf{k} \cdot \mathbf{p}) \bl( \frac{1}{E} C_{ii} (\mathbf{p}) \br) & = & 0 \ , \label{appA9}
\end{eqnarray}
and therefore the evolution equations for the macroscopic quantities $\overline{\rho}_i$, $\delta_i$, $\theta_i$ cannot have contributions from $ii$ self-scattering. This is a great simplification which will allow us to ignore the self-scattering collision terms in the perturbation equations for $\delta_i$ and $\theta_i$. However, they do not vanish in the equations for the distribution functions.

\subsection{The IDM equations}
The macroscopic IDM perturbation equations are relatively simple. Taking the appropriate moments of \Eq{appA3} and expanding to first order in the perturbations one obtains (see \cite{Ma:1995ey,Cyr-Racine:2015ihg}):
\begin{eqnarray}
\dot{\delta}_\idm + \theta_\idm - 3 \dot\phi & = & 0 \ , \label{appA10} \\
\dot\theta_\idm - c_\idm^2 k^2 \delta_\idm + \mH \theta_\idm - k^2 \psi & = & \frac{a (1 + \psi)}{M \rho_\idm} g_\idm \int \frac{\text d^3 \mathbf{p}}{(2 \pi)^3} \,\, (i \mathbf{k} \cdot \mathbf{p}) C_{\idm-\dr} (\mathbf{p}) \ , \label{appA11}
\end{eqnarray}
where from now on we denote $\overline{\rho}$ by $\rho$, and where $c_\idm^2 = \frac{\dot P_\idm}{\dot\rho_\idm}$ is the speed of sound of the IDM. As explained in the previous subsection, energy-momentum conservation sets to zero any contributions of the $C_{\idm-\idm}$ collision term to $\delta$ and $\theta$. Note that in the non-relativistic limit the kinetic energy of the IDM particles is negligible compared with the mass so that $\rho_\idm \approx M n_\idm$, and therefore \Eq{appA10} becomes equivalent to IDM particle number conservation. There is no contribution from the collision term $C_{\idm-\dr}$ because the scattering preserves IDM particle number in the non-relativistic limit. There would be a contribution from $C_{\idm-\dr}$ if the IDM-DR interactions were to significantly heat up the IDM so that $\rho_\idm$ contains a non-negligible kinetic energy contribution. We do not consider such a case in this paper. 
Additional equations determining the gravitational potential perturbations $\phi$ and $\psi$ follow from Einstein's equation \cite{Ma:1995ey}. 

\subsection{The DR equations}
For the DR equations, a more careful treatment of the $\Psi_\dr$ fluctuations is required. We refer to \cite{Cyr-Racine:2015ihg} for a detailed description of the DR perturbation equations, here we just quote the results which are needed for our discussion.

We assume that the DR fluid is approximately in thermal equilibrium so that the zeroth-order $\f0_\dr$ is given by the Bose-Einstein or a Fermi-Dirac distribution function:
\begin{equation}\label{appA12}
\f0_\dr (q) = \frac{1}{e^{q/(aT_\dr)} \mp 1} \ , \quad T_\dr \propto a^{-1} \ .
\end{equation}
Note that this distribution function is time-independent when expressed in terms of the co-moving momentum $q$. The perturbations $\Psi(q, k, \mu \equiv \hat{k} \cdot \hat{q}, \eta)$ are small, local deviations from this homogeneous and isotropic thermal equilibrium.

Following \cite{Ma:1995ey} (\cite{Cyr-Racine:2015ihg} use $F_l$ instead of $\Psi_l$) we expand the $\mu$-dependence of the perturbations $\Psi_\dr(q, k, \mu, \eta)$ in Legendre Polynomials $P_l(\mu)$
\begin{equation}\label{appA13}
\Psi_\dr(q, k, \mu, \eta) = \sum\limits_{l=0}^\infty (-i)^l (2l+1) \Psi_l(q,k,\eta) P_l(\mu) \ .
\end{equation}

To linear order in the perturbations, the full Boltzmann equations (\Eq{appA4}) simplify into a coupled system of equations for the $\Psi_l$
\begin{eqnarray}\label{appA14}
\f0_\dr \bl[ \frac{\partial \Psi_l}{\partial \tau} + k \bl( \frac{l+1}{2l+1}\Psi_{l+1} - \frac{l}{2l+1}\Psi_{l-1} \br) \br. & & \nn\\
 \bl. + \dlnf0 \bl( \frac{\partial \phi}{\partial \tau}\delta_{l0} + \frac{k}{3} \psi \delta_{l1} \br) \br] & = & \frac{a^2}{q} \frac{(-i)^{-l}}{2} \int\limits_{-1}^1 \,\, \text d\mu P_l(\mu) C^{(1)}(\mathbf{q}/a) \ ,
\end{eqnarray}
where $\delta_{li}$ is the Kronecker delta and  $C^{(1)}(\mathbf{q}/a) = C^{(1)}_{\dr-\idm}(\mathbf{q}/a) + C^{(1)}_{\dr-\dr}(\mathbf{q}/a)$ is the first-order DR collision term.

We can re-express the macroscopic DR fluid variables \Eqsto{appA5}{appA7} in terms of the $\Psi_l$
\begin{eqnarray}
\rho_\dr & = & a^{-4} \frac{g_\dr}{2 \pi^2} \int\limits_0^\infty \text dq \,\, q^3 \f0_\dr \ , \label{appA15} \\
F_l & \equiv & \frac{g_\dr}{2 \pi^2} \bl( \rho_\dr a^4 \br)^{-1} \ \int\limits_0^\infty \text dq \,\, q^3 \f0_\dr \ \Psi_{l} \ , \label{appA16} \\
\delta_\dr & = & F_0 \ , \label{appA17} \\
\theta_\dr & = & \frac{3}{4}k F_1 \ . \label{appA18} 
\end{eqnarray}
With these expressions in mind, one can integrate \Eq{appA14} to obtain the cosmological linear perturbations equations for the DR.

\subsection{The DR-DM collision term}
We now discuss the IDM-DR collision term. Using results for $C_{\dr-\idm}$ from \cite{Cyr-Racine:2015ihg}, \Eq{appA14} can be written as
\begin{eqnarray}\label{appA19}
\f0_\dr \bl[ \frac{\partial \Psi_l}{\partial \tau} \br. \! & + & \!\! \bl. k \bl( \frac{l+1}{2l+1}\Psi_{l+1} - \frac{l}{2l+1}\Psi_{l-1} \br) + \dlnf0 \bl( \frac{\partial \phi}{\partial \tau}\delta_{l0} + \frac{k}{3} \psi \delta_{l1} \br) \br] \nn\\
& = & -\frac{a \rho_\idm}{16 \pi M^3} \f0_\dr \bl[ \Delta_l(q/a) \Psi_l + \delta_{l1} \Delta_1(q/a) \frac{\theta_\idm}{3k} \dlnf0 \br] -a \f0_\dr \Lambda_l(q/a) \Psi_l \ ,
\end{eqnarray}
where $\Lambda_l(q/a)$ accounts for the DR-DR self-interactions and $\Delta_l(p=q/a)$ describe DR-IDM scattering
\begin{eqnarray}
\Delta_l(p=q/a)  & \equiv & \frac{1}{2} \int\limits_{-1}^1 d\tilde{\mu} \,\, (1 -P_l(\tilde{\mu}))\vert \overline{\mathcal{M}} \vert^2_{\dr-\idm} \Big\vert_{\substack{t = 2p^2(\tilde{\mu}-1) \\ s = M^2 + 2 p M}} \ . \label{appA20}
\end{eqnarray}
Here $\tilde{\mu} \equiv \hat{q}_\mathrm{in} \cdot \hat{q}_\mathrm{out}$ is the cosine of the angle between the ingoing and the outgoing DR particles in the DR-IDM collisions. Note that the contribution from DR-IDM collisions vanishes for $l=0$. This is because in the non-relativistic limit of the IDM the scattering of DR off of IDM does not change the energy of the DR. As already mentioned earlier, energy-momentum conservation, \Eqs{appA8}{appA9}, implies that
\begin{equation}\label{appA21}
\int \text d q \,\, q^3 \f0_\dr \Lambda_{0,1}(q/a) \Psi_{0,1} = 0 \ .
\end{equation}
Note however that $\Lambda_{0,1}(q/a)$ do not vanish, and other moments of $\Lambda_{0,1} \Psi_{0,1}$ from the ones appearing in \Eq{appA21} are not zero.

Using \Eqsto{appA15}{appA18} in \Eq{appA19} one obtains the Boltzmann hierarchy of linear perturbation equations for the DR fluid
\begin{eqnarray}
\dot\delta_\dr + \frac{4}{3} \theta_\dr - 4 \dot\phi & = & 0 \ , \label{appA22} \\
\dot\theta_\dr + k^2 \bl( \frac12 F_2 - \frac{1}{4} \delta_\dr \br)	 - k^2 \psi & = & \bl( \frac{3}{4}\frac{\rho_\idm}{\rho_\dr} \br) \Bigg[ \frac{a^{-3} g_\dr}{24 \pi^3 M^3} \int \text dq \,\, q^3 \f0_\dr \Delta_1 \nn\\
& & \times \bl( -\frac{1}{4} \dlnf0 \theta_\idm - \frac{3}{4} k \Psi_1 \br) \Bigg] \ , \label{appA23} \\
\dot{F}_l + \frac{k}{2l+1} \bl( (l+1) F_{l+1} - l F_{l-1} \br) & = & - \bl( \frac{3}{4}\frac{\rho_\idm}{\rho_\dr} \br) \bl[ \frac{a^{-3} g_\dr}{24 \pi^3 M^3} \int \text dq \,\, q^3 \f0_\dr \Delta_l \Psi_l \br]  \nn\\
& & - \frac{a^{-3} g_\dr}{2 \pi^2 \rho_\dr} \int \text dq \,\, q^3 \f0_\dr \Lambda_l(q/a) \Psi_l \ , \quad\quad l \geq 2 \ . \label{appA24}
\end{eqnarray}
In general, $\Delta_l(q/a)$ as well as $\Psi_l$ depend on $q$ so that the integrals on the r.h.s. of \Eqsto{appA23}{appA24} cannot be re-expressed in terms of the moments $F_l$ (\Eq{appA16}).

In order to compare the IDM with the DR equations it is convenient to write the collision term in \Eq{appA11} in terms of $C_{\dr-\idm}(q/a)$ using momentum conservation (\cite{Dodelson:2003ft})
\begin{equation}\label{appA25}
g_\idm \int \frac{\text d^3 \mathbf{p}_\idm}{(2 \pi)^3} \,\, \frac{\mathbf{p}_\idm}{M} C_{\idm-\dr}(\mathbf{p}_\idm) = - g_\dr \int \frac{\text d^3 \mathbf{p}_\dr}{(2 \pi)^3} \,\, \hat{p}_\dr C_{\dr-\idm}(\mathbf{p}_\dr) \ .
\end{equation}
Then the $\theta_\idm$ equation becomes (\cite{Cyr-Racine:2015ihg})
\begin{eqnarray}\label{appA26}
\dot{\theta}_\idm - c_\idm^2 k^2 \delta_\idm + \mathcal{H} \theta_\idm - k^2 \psi & = & \Bigg[ \frac{a^{-3} g_\dr}{24 \pi^3 M^3} \int \text dq \,\, q^3 \f0_\dr \Delta_1 \nn\\
& & \times \bl( \frac{3}{4} k \Psi_1 - \bl( -\frac{1}{4} \dlnf0 \br)\theta_\idm \br) \Bigg] \ .
\end{eqnarray}
Comparing \Eqs{appA23}{appA26} we see that the interaction terms in the macroscopic equations for the IDM and the DR are proportional to each other with proportionality factor $-{3 \rho_\idm}/{4 \rho_\dr}$.

\subsection{The 3/4 factor}

The ${3 \rho_\idm}/{4 \rho_\dr}$ factor relating the two collision terms can also be obtained directly from conservation of the stress-energy-tensor of the DM and DR fluids in their macroscopic description (\ie without going into the microscopic details of a Boltzmann equation).

Following for example \cite{Uzan:1998mc,Malik:2008im}, consider two fluids $A$ and $B$ which interact with one another respecting overall energy-momentum conservation, $\nabla_\mu \bl ( T^{\mu\nu}_A + T^{\mu\nu}_B \br) = 0$. This implies that if $\nabla_\mu T^{\mu\nu}_A = Q^\nu$ then $\nabla_\mu T^{\mu\nu}_B = -Q^\nu$ where $Q$ is the force $B$ exerts on $A$ (Newton's third law).
From these conservation equations for $T^{\mu\nu}_{A,B}$ one derives the Euler equations for the perturbations $\theta_{A,B}$ (see \Eq{appA7}). 
When re-expressed in terms of the $\theta_{A,B}$, Newton's 3rd law becomes that the force term in the equation for $\theta_A$ that accounts for its interactions with $B$ is equal and opposite to its counterpart in the equation for $\theta_B$ times the proportionality factor $({\rho_B + P_B})/({\rho_A + P_A})$, independent of any details of the interactions $Q$. This factor reduces to the familiar $3{\rho_\idm}/4{\rho_\dr}$ for the case of matter-radiation interactions that we are interested in.

We note that (v3) and earlier versions of \cite{Cyr-Racine:2015ihg} found a proportionality factor which depends on the scaling of the IDM-DR interaction term $\Delta_l(q/a)$ with $q$. Naively, this would seem to contradict momentum conservation, which as we saw requires a proportionality factor that is independent of the details of the interaction. The disagreement comes from a non-standard definition of $\theta_{dr}^{ETHOS}$ used in (v3) of \cite{Cyr-Racine:2015ihg}. Using our notation, \cite{Cyr-Racine:2015ihg} defined
\begin{equation}
\theta_\dr^{ETHOS} = -3k \frac{g_\dr}{2 \pi^2} \bl( \rho_\dr a^4 \br)^{-1} \ 
\int\limits_0^\infty \text dq \,\, q^3 \f0_\dr \bl( \dlnf0 \br)^{-1} \ \Psi_{1} \label{appA27} 
\end{equation}
instead of the conventional 
\begin{equation}
\theta_\dr = \frac{3}{4}k \frac{g_\dr}{2 \pi^2} \bl( \rho_\dr a^4 \br)^{-1} \ 
\int\limits_0^\infty \text dq \,\, q^3 \f0_\dr \ \Psi_{1} \ . \label{appA28} 
\end{equation}
The important difference is that with the $( \dlnf0)^{-1}$ factor under the $q$-integral $\theta_\dr^{ETHOS}$ is not simply related to the momentum density of the DR fluid. Thus momentum conservation is rather tricky to understand using the variables used in \cite{Cyr-Racine:2015ihg}. For example, in the ETHOS formalism the DR-DR self-interaction term (in the ETHOS equivalent of our \Eq{appA21}) cannot be argued to vanish using momentum conservation, and the ETHOS equivalent of our \Eq{appA23} contains an inconvenient additional DR-DR collision term. The difference between the two definitions of $\theta_\dr$ vanishes for ``locally thermal" distribution functions (defined below). This might tempt one to expect that the perturbation equations obtained in the two formalisms should be the same when frequent DR-DR collisions lead to local thermality. However this is incorrect: the IDM-DR collisions drive the DR distribution function away from thermality and the DR-DR collision term does not vanish even when expanding about thermal distribution functions. 

\subsection{Simplifying the interaction terms: thermal approximation}
In the general case the r.h.s. of \Eqs{appA23}{appA24} as well as \Eq{appA26} cannot be rewritten in terms of the moments $\theta_\dr$ or $F_l$. Thus the differential equations for the $F_l$ do not form a closed system of equations and cannot be solved. Instead one has to go back to the un-integrated  Boltzmann equations \Eq{appA19} and solve for the full $q$-dependence of the perturbations $\Psi_l(q/a)$. Fortunately, in our case of interest this is not necessary because we can greatly simplify the collision terms by making the physically motivated assumption of approximate ``local thermality" of the dark radiation fluid.

To better understand the physical situation it is useful to compare the collision rates of the IDM-DR and DR-DR interactions. The rate of momentum transfer in the IDM-DR collisions scales as $\alpha^2 T_\dr^2/M$  whereas the rate of DR-DR collisions scales as $\alpha^2 T_\dr$. Since $T_\dr \ll M$ the dark radiation fluid self-scatters many times between IDM-DR collisions. These self-scatters efficiently redistribute energy and momentum among the DR particles and allow the DR fluid to attain local thermal equilibrium. Thus to a very good approximation we may assume that every time an IDM-DR collision occurs the radiation particle is drawn from a thermal distribution. 

In the above argument the term ``local thermal equilibrium" means that any point in space-time one can boost to a frame such that the distribution function is thermal. In Fourier space and for small fluctuations this means that the temperature of the distribution may depend on the direction and magnitude of the wave vector $\mathbf{k}$ and time so that we may write
\begin{equation}\label{appA29}
f_\dr \bl( \frac{\mathbf{q}}{a}, \mathbf{k},\eta \br) = \frac{1}{e^{q/(aT(k,\mu,\eta))} \mp 1} \equiv \frac{1}{e^{q/\bl[(aT_\dr) (1 + \frac{1}{4}\Theta(k,\mu,\eta)\br]} \mp 1} \approx \f0_\dr(q) \bl( 1 - \frac{1}{4} \dlnf0 \Theta(k,\mu, \eta) \br) \ ,
\end{equation}
where $\mu \equiv \hat{k} \cdot \hat{q})$. Comparing with \Eq{appA1} we see that the perturbations are
\begin{equation}\label{appA30}
\Psi_\dr ( q,k,\mu, \eta) =  - \frac{1}{4} \dlnf0 \Theta(k,\mu, \eta) \ ,
\end{equation}
where the $q$ dependence is isolated in the prefactor $\dlnf0$. Expanding both sides in Legendre polynomials we obtain 
\begin{equation}\label{appA31}
\Psi_l ( q, k,\eta) =  - \frac{1}{4} \dlnf0 \Theta_l(k,\eta) \ ,
\end{equation}
where again it is important to note that the $\Theta_l$ are $q$-independent (the authors of \cite{Cyr-Racine:2015ihg} use $\nu_l$ instead of $\Theta_l$).

Using this in \Eqsto{appA16}{appA18} we find that, after integrating by parts:
\begin{equation}\label{appA32}
\delta_\dr = \Theta_0 \ , \quad \theta_\dr = \frac{3}{4}k \Theta_1 \ ,  \quad F_{l} = \Theta_{l} \ .
\end{equation}
Thus the r.h.s. of \Eqsto{appA23}{appA24} and \Eq{appA26} can now be rewritten in terms of the macroscopic quantities of the DR fluid
\begin{eqnarray}
\dot{\delta}_\idm + \theta_\idm - 3 \dot\phi & = & 0 \ , \label{appA33} \\
\dot\theta_\idm - c_\idm^2 k^2 \delta_\idm + \mH \theta_\idm - k^2 \psi & = & \gamma_1(a) \bl( \theta_\dr - \theta_\idm \br) \ ; \label{appA34}
\end{eqnarray}
\begin{eqnarray}
\dot\delta_\dr + \frac{4}{3} \theta_\dr - 4 \dot\phi & = & 0 \ , \label{appA35} \\
\dot\theta_\dr + k^2 \bl(\frac12 F_2 - \frac{1}{4} \delta_\dr \br) - k^2 \psi & = & \bl( \frac{3}{4}\frac{\rho_\idm}{\rho_\dr} \br) \gamma_1(a) \bl( \theta_\idm - \theta_\dr \br) \ , \label{appA36} \\
\dot{F}_l + \frac{k}{2l+1} \bl( (l+1) F_{l+1} - l F_{l-1} \br) & = & - \bl[\bl( \frac{3}{4}\frac{\rho_\idm}{\rho_\dr} \br)\gamma_l(a) + \lambda_l(a) \br] F_l \ , \ l \geq 2 \ ; \label{appA37}
\end{eqnarray}
\begin{eqnarray}
\gamma_l(a) & \equiv & \frac{a^{-3} g_\dr}{24 \pi^3 M^3} \int \text dq \,\, q^3 \f0_\dr \bl( -\frac{1}{4} \dlnf0 \br) \Delta_l \ , \label{appA38} \\
\lambda_l(a) & \equiv & \frac{a^{-3} g_\dr}{2 \pi^2 \rho_\dr} \int \text dq \,\, q^3 \f0_\dr \bl( -\frac{1}{4} \dlnf0 \br) \Lambda_l(q/a) \label{appA39} \ .
\end{eqnarray}
We recall that energy and momentum conservation imply that $\lambda_{0,1}=0$. Finally, for a fluid which is in local thermal equilibrium all moments of the Boltzmann hierarchy which are not conserved (i.e. all except the energy and momentum densities) are driven to zero (``perfect fluid"). Therefore the DR equations simplify even further
\begin{eqnarray}
\dot\delta_\dr + \frac{4}{3} \theta_\dr - 4 \dot\phi & = & 0 \ , \label{appA40} \\
\dot\theta_\dr - \frac{k^2}{4} \delta_\dr - k^2 \psi & = & \bl( \frac{3}{4}\frac{\rho_\idm}{\rho_\dr} \br) \mG(a) \bl( \theta_\idm - \theta_\dr \br) \ ,\label{appA41}
\end{eqnarray}
where we have renamed $\mG(a) \equiv \gamma_1(a)$.
For particle physics models that realize such a scenario, see \cite{Buen-Abad:2015ova,Lesgourgues:2015wza,Ko:2016uft,Ko:2016fcd,Ko:2017uyb}.

Note that one can easily generalize the above to a scattering matrix element which gives the momentum dependence $\Delta_l(q/a) = d_l (q/a)^n$ (\cite{Cyr-Racine:2015ihg}), we find
\begin{equation}\label{appA42}
\gamma_l(a) = g_\dr d_l \frac{\bl( 1+\frac{n}{4} \br) \xi(n) \Gamma(4+n) \zeta(4+n)}{24 \pi^3 M^3} \ (a T_\dr)^{4+n} a^{-n-3} \ ,
\end{equation}
where $\xi(n) = 1 \ (1-2^{-3-n})$ for bosons (fermions).

\section{Analytic solutions and the shape of the MPS}
\label{appB}

Our goal in this appendix is to obtain analytic expressions for the suppression:
\begin{equation}\label{appB1}
	S \equiv \frac{\delta_{\mathrm{clumping \ DM}}}{\delta_\cdm^{\LC}} \ ,
\end{equation}
where \textit{``clumping DM"} refers to the IDM in the WI limit or the CDM in the DP limit. This suppression then yields our model's prediction for the shape of the MPS compared to that of \LC. In order to do this we need to rewrite the perturbation equations in more appropriate variables.

Defining $a_\eq \equiv \ora /\oma$, $\a \equiv a / a_\eq$, $\mH \equiv a H(a)$, $\mG \equiv a \Gamma(a) = a^{-1} \g0$, and $\ok(\a) \equiv k/\mH$, the \Eqs{didm_2}{ddr_2} for the cosmological perturbations (in the conformal Newtonian Gauge, see \cite{Ma:1995ey}) of $\delta_\idm$ and $\delta_\dr$ as a function of $\a$ are:

\begin{eqnarray}
	\a^2 \delta_\idm'' + \bl( 2 + \frac{\a \mH'}{\mH} + \frac{\mG}{\mH} \br) \a \delta_\idm' & = & -\ok^2 \psi + 3\a^2 \phi'' + 3\bl( 2 + \frac{\a \mH'}{\mH} \br)\a \phi' \nn\\
	& & + \frac{\mG}{\mH} \frac{3}{4} \a \delta_\dr' \label{appB2} \\
	\a^2 \delta_\dr'' + \bl( 1 + \frac{\a \mH'}{\mH} + \frac{R \mG}{\mH} \br) \a \delta_\dr' + \frac{\ok^2}{3}\delta_\dr & = & \frac{4}{3} \bl[ -\ok^2 \psi + 3\a^2 \phi'' + 3\bl( 1 + \frac{\a \mH'}{\mH} \br)\a \phi' \br. \nonumber \\
	& & \quad \ \bl. +\frac{R \mG}{\mH} \a \delta_\idm' \br] \ , \label{appB3}
\end{eqnarray}
while \Eqs{didm_3}{del_1} are:
\begin{eqnarray}
	\a^2 \delta_\idm'' + \bl( 2-3 \csp^2 + \frac{\a \mH'}{\mH} \br) \a \delta_\idm' + \ok^2 \csp^2 \delta_\idm && \nonumber\\
	= - \ok^2 \psi + 3 \a^2 \phi'' + 3 \bl( 2-3 \csp^2 + \frac{\a \mH'}{\mH} \br)\a \phi' \!\! & + & \!\! 3 \csp^2 \bl[ \a^2 \Delta'' + \bl( 1 + \frac{\a \mH'}{\mH} \br)\a \Delta' + \frac{\ok^2}{3}\Delta \br] \label{appB4} \\
	3 \csp^2 \bl[ \a^2 \Delta'' + \bl( 1 + \frac{\a \mH'}{\mH} \br) \a \Delta' + \frac{\ok^2}{3} \Delta \br] + \frac{\mG}{\mH} \a \Delta' & = & 3 \csp^2 \bl[ 3 \a \phi' - \a \delta_\idm' + \frac{\ok^2}{3}\delta_\idm \br] \ . \label{appB5}
\end{eqnarray}
All the derivatives are with respect to $\a$.

The metric perturbations $\psi$ and $\phi$, meanwhile, obey:
\begin{eqnarray}
	\ok^2 \phi + 3\bl( \a \phi' + \psi \br) & = & -\frac{3}{2}\bl( \frac{\h0}{h} \br)^2 \sum_{i} \omega_i \bl(\frac{\ora}{\oma}\br)^{-(1+3 w_i)} \frac{\a^{-(1+3 w_i)}}{\mH^2} \delta_i \label{appB6} \\
	\a^2 \phi'' + \bl( 3 + \frac{\a \mH'}{\mH} \br)\a \phi' + \a\psi' \!\!\! & + & \!\!\! \bl( 1 + \frac{\a \mH'}{\mH} \br)\psi 
	+ \frac{\ok^2}{3}\bl( \phi - \psi \br) \nn\\
	& = & \frac{3}{2}\bl( \frac{\h0}{h} \br)^2 \sum_{i} \omega_i \bl(\frac{\ora}{\oma}\br)^{-(1+3 w_i)} c_{si}^2 \frac{\a^{-(1+3 w_i)}}{\mH^2} \delta_i \label{appB7}\ ,
\end{eqnarray}

where $w_i \equiv P_i / \rho_i$, and $c_{si}$ is the speed of sound, for each component $i$ of the Universe.

We are interested in finding solutions to \Eqs{appB2}{appB3} for $\a \ll \a_\Lambda$, and ignoring the baryons and anisotropic stress.

Under these conditions
\begin{equation}\label{appB8}
	\mH^2 = \bl( \frac{\h0}{h} \br)^2\frac{\oma^2}{\ora} \bl(\a^{-1} + \a^{-2} + \bl(\frac{a_\eq}{a_\Lambda}\br)^3 \a^2 \br) \approx \bl( \frac{\h0}{h} \br)^2\frac{\oma^2}{\ora} \frac{\a + 1}{\a^2} \ ,
\end{equation}
\begin{equation}\label{appB9}
	\ok \approx \ok_\eq \frac{\sqrt{2} \a}{\sqrt{1 + \a}} \ .
\end{equation}
\Eqs{appB2}{appB3} can be written as:
\begin{eqnarray}
\a^2 \delta_\idm'' + \bl( \frac{1 + \frac{3}{2}\a}{1+\a} + \frac{g}{\sqrt{1+\a}} \br) \a \delta_\idm' & = & -\ok^2 \psi + 3\a^2 \phi'' + 3\bl( \frac{1 + \frac{3}{2}\a}{1+\a} \br)\a \phi' \nonumber \\
& & + \frac{g}{\sqrt{1+\a}} \frac{3}{4} \a \delta_\dr' \label{appB10} \\
\a^2 \delta_\dr'' + \bl( \frac{\a/2}{1+\a} + \frac{Rg}{\sqrt{1+\a}} \br) \a \delta_\dr' + \frac{\ok^2}{3}\delta_\dr & = & \frac{4}{3} \bl[ -\ok^2 \psi + 3\a^2 \phi'' + 3\bl( \frac{\a/2}{1+\a} \br)\a \phi' \br. \nonumber \\
& & \bl. + \frac{Rg}{\sqrt{1+\a}}\a \delta_\idm' \br] \ , \label{appB11}
\end{eqnarray}
\begin{equation}
\mathrm{where} \quad g \equiv \bl. \frac{\sqrt{2} \mG}{\mH}\right\vert_\eq \approx \frac{h \g0}{\ora^{1/2} \h0} \approx 10^2 \frac{\g0}{\h0} \ . \label{appB12}
\end{equation}

And \Eqs{appB4}{appB5} as:
\begin{eqnarray}
\a^2 \delta_\idm'' + \bl( \frac{1+\frac{3}{2}\a}{1+\a}-3 \csp^2 \br) \a \delta_\idm' + \ok^2 \csp^2 \delta_\idm & = & - \ok^2 \psi + 3 \a^2 \phi'' + 3 \bl( \frac{1+\frac{3}{2}\a}{1+\a}-3 \csp^2 \br)\a \phi' \nn\\
+ 3 \csp^2 \bigg[ \a^2 \Delta'' \!\!\! & + & \!\!\! \bl. \bl( \frac{\a/2}{1+\a} \br)\a \Delta' + \frac{\ok^2}{3}\Delta \br] \label{appB13} \\
3 \csp^2 \bl[ \a^2 \Delta'' + \bl( \frac{\a/2}{1+\a} \br) \a \Delta' + \frac{\ok^2}{3} \Delta \br] + \bl( \frac{g}{\sqrt{1+\a}} \br) \a \Delta' & = & 3 \csp^2 \bl[ 3 \a \phi' - \a \delta_\idm' + \frac{\ok^2}{3}\delta_\idm \br] \ . \label{appB14}
\end{eqnarray}

We are interested in analytic solutions for these equations in order to describe the shape of the MPS for the WI, DP limits, for the modes of interest: $k \gg k_\eq$. Because these modes are already well inside the Hubble radius early during RD, we follow Weinberg's method (\cite{Weinberg:2002kg,Weinberg:2008zzc}) of obtaining the solutions: we take the $\a \ll 1$ limit for the above equations and solve; then we take the $\ok \gg 1$ limit and solve for the slow modes (which are the growing ones), and subsequently match the results.

\subsection{Weakly Interacting}
In this limit $g \ll 1$. Let us first solve the equations for the early Radiation Domination era, when $\a \ll 1$. In this case the right hand sides of \Eqs{appB6}{appB7} are proportional to each other (because only the radiation components contribute, and $c_{si}^2 = 1/3$). Therefore, combining these equations yields:
\begin{equation}
\a^2 \psi'' + 4 \a \psi' + \bl( \frac{2 \ok_\eq^2}{3} \br)\a^2 \psi \approx 0 \label{appB15} \ ,
\end{equation}
where, in ignoring the anisotropic stress, we have taken $\psi \approx \phi$.

A more useful variable is $y \equiv \a/\a_k$, where $\a_k$ is defined by $\ok(\a_k) = 1$: the time when the $k$ mode enters the Hubble radius. For $k \gg k_\eq$, $\a_k^{-1} \approx \sqrt{2} \ok_\eq$.

The solution to \Eq{appB15} is then
\begin{equation} \label{appB16}
\psi_\mathrm{sol}(y) \approx 3 \frac{\sin\bl(\frac{y}{\sqrt{3}}\br) - \frac{y}{\sqrt{3}} \cos\bl(\frac{y}{\sqrt{3}} \br)}{\bl( y/\sqrt{3} \br)^3} \ ,
\end{equation}
for the adiabatic initial conditions $\psi(0) = 1$ and $\psi'(0) = 0$.

For $\a \ll 1$ then $R \ll 1$. Using this, and dropping $Rg \ll g \ll 1$, \Eqs{appB10}{appB11} become:
\begin{eqnarray}
	y^2 \delta_\idm'' + \bl( 1 + g \br) y \delta_\idm' & \approx & 3y^2 \psi'' + 3 y \psi' -y^2 \psi + g\frac{3}{4} y \delta_\dr' \ , \label{appB17} \\
	\delta_\dr'' + \frac{1}{3}\delta_\dr & \approx & \frac{4}{3} \bl( 3 \psi'' - \psi \br) \ . \label{appB18}
\end{eqnarray}
With \Eq{appB16} we can solve \Eqs{appB17}{appB18}. In particular, the solution for $\delta_\idm$ that satisfies its adiabatic initial conditions is:
\begin{equation}\label{appB19}
	\delta_\idm \approx - \frac{3}{2} + \int\limits_{0}^{y} dx \ \bl( \frac{1-(y/x)^{-g}}{g x} \br) \bl( 3 x^2 \psi'' + 3x \psi' - x^2 \psi + \frac{3}{4} g x \delta_\dr' \br) \ ,
\end{equation}
$\bl( \frac{1-(y/x)^{-g}}{g x} \br)$ being the Green's function of the $\delta_\idm$ differential equation. This integral is doable, but because we will be interested in modes well inside the Hubble radius, we take $y \gg 1$ (which can always be done for any $\a \ll 1$ given a sufficiently large $k$). This yields:
\begin{equation}\label{appB20}
	\delta_\idm \xrightarrow{y \gg 1} \frac{9}{2} \bl( -\frac{2}{g} +(2+g)\cos (g\pi/2)\Gamma(g) \bl( \frac{y}{\sqrt{3}} \br)^{-g} \br) \ ,
\end{equation}
which for $g=0$ (i.e. $\delta_\cdm$ in \LC) gives $-\frac{9}{2} \bl( -1 + 2 \gamma_E - \log 3 + 2\log y \br)$, where $\gamma_E$ is the Euler-Mascheroni constant (see \cite{Weinberg:2002kg,Weinberg:2008zzc}).

Let us now proceed to study the slow modes deep inside the Hubble radius, i.e. the case with $\dot\delta \sim \mH \delta$ and $\ok \gg 1$. It can be proven (see \cite{Weinberg:2002kg,Weinberg:2008zzc} for details) that for these slow modes $\delta_\ra \sim \ok^{-2} \delta_\ma$ and $\rho_\ra \delta_\ra \ll \rho_\ma \delta_\ma$. With this in mind, we can show that the only contribution to the right hand side of \Eq{appB10} comes from the IDM:
\begin{equation}\label{appB21}
	\ok^2 \psi \approx - \frac{\frac{3}{2}\a}{1 + \a} \delta_\idm \ .
\end{equation}
It can be shown that for the slow modes $\psi$ is a constant and therefore, ignoring $\delta_\dr' \sim \ok^{-2} \delta_\idm'$, \Eq{appB10} reduces to a modified version of the M\'{e}sz\'{a}ros equation (\cite{Meszaros:1974tb}):
\begin{equation}\label{appB22}
	\a^2 \delta_\idm'' + \bl( \frac{1 + \frac{3}{2}\a}{1+\a}+\frac{g}{\sqrt{1+\a}} \br) \a \delta_\idm' -\frac{\frac{3}{2}\a}{1+\a}\delta_\idm \approx 0 \ .
\end{equation}
The two independent solutions to this equation are:
\begin{eqnarray}
\delta_1 & \equiv & \bl( \frac{\a}{\sqrt{1+\a}-1} \br)^g \leftidx{_2}F_1\bl[-1+\frac{g}{2},\frac{3}{2}+\frac{g}{2};1+g;-\a\br] \ , \label{appB23}\\
\delta_2 & \equiv & \bl( \frac{1}{\sqrt{1+\a}-1} \br)^g \leftidx{_2}F_1\bl[-1-\frac{g}{2},\frac{3}{2}-\frac{g}{2};1-g;-\a\br] \ . \label{appB24}
\end{eqnarray}
Note that for $\a \ll 1$ these solutions give:
\begin{eqnarray}
\delta_1(\a \ll 1) & \sim & 2^g \ , \label{appB25}\\
\delta_2(\a \ll 1) & \sim & 2^g \a^{-g} = 2^g \a_k^{-g} y^{-g} \ . \label{appB26}
\end{eqnarray}
Applying Weinberg's matching method to our WI model, we find the linear combination of $\delta_1$ and $\delta_2$ that has the same $y$ dependence as \Eq{appB20}:
\begin{equation}\label{appB27}
	\delta_\idm \approx \bl( -\frac{9}{2^g g} \br) \delta_1 + \bl( \frac{3^{2+g/2} \a_k^g}{2^{1+g}} (2+g) \cos(g \pi/2) \Gamma(g)  \br) \delta_2 \ .
\end{equation}
It can be shown that for $g=0$ this reduces to the well-known solution to the M\'{e}sz\'{a}ros equation.

Because we are interested in the shape of the MPS at late times, it is useful to take the $\a \gg 1$ limit and compare the result to $\LC+\dN$: the $g=0$ case. This yields:

\begin{equation}\label{appB28}
	S(\a)_\mathrm{WI} \xrightarrow{\a \gg 1} \frac{4-6g+2g^2 -\bl( \frac{0.36}{k \eta_\eq} \br)^g(2+g)\cos(g \pi/2)\Gamma(3+g)}{g(4-5g^2+g^4)(-1.9+\log k \eta_\eq)} \ ,
\end{equation}

where we have used $\a_k^{-1} \approx \sqrt{2} \ok_\eq \approx \frac{k \eta_\eq}{2(\sqrt{2}-1)}$, true for $k \gg k_\eq$. As a series in $g$ this expression can be rewritten as
\begin{equation}\label{appB29}
	S(\a \gg 1)_\mathrm{WI} \approx \frac{\frac{1-\bl( \frac{0.36}{k \eta_\eq} \br)^g}{g} - \bl( 1.5+1.42 \bl( \frac{0.36}{k \eta_\eq} \br)^g \br) + g \bl( 1.75 - 1.1 \bl( \frac{0.36}{k \eta_\eq} \br)^g \br) }{-1.9 + \log k \eta_\eq} \ .
\end{equation}

For large $\log k \eta_\eq$ but $g$ small enough that $g \log(k\eta_\eq/0.36) \ll 1$, we find that the above expression goes like:

\begin{equation}\label{appB30}
	S(\a \gg 1)_\mathrm{WI} \sim 1 - \frac{g}{2} \log k \eta_\eq \ .
\end{equation}

\Eq{appB29} then allows us to compare the MPS of the WI model and the \LCs case, which gives a suppression of the form:
\begin{equation}\label{appB31}
	\frac{P(k)_{\mathrm{WI}}}{P(k)_{\LC+\dN}} = \frac{\langle \delta_\ma \delta_\ma \rangle}{\langle \delta_\ma \delta_\ma \rangle\vert_{g=0}} \approx \frac{\langle \delta_\idm \delta_\idm \rangle}{\langle \delta_\cdm \delta_\cdm \rangle\vert_{g=0}} \approx \bl[ S(\a \gg 1)_\mathrm{WI} \br]^2 \ .
\end{equation}

\subsection{Dark Plasma}
In this subsection we obtain analytic solutions to the DP limit of our model, deriving the results from \cite{Chacko:2016kgg} in a more precise manner. In this limit $\g0 \gg \h0$ and thus $\Delta = 0$; and there are two DM components: CDM and IDM. \Eq{appB13} becomes:
\begin{equation}\label{appB32}
	\a^2 \delta_\idm'' + \bl( \frac{1+\frac{3}{2}\a}{1+\a}-3 \csp^2 \br) \a \delta_\idm' + \ok^2 \csp^2 \delta_\idm = 3 \a^2 \psi'' + 3 \bl( \frac{1+\frac{3}{2}\a}{1+\a}-3 \csp^2 \br)\a \psi' - \ok^2 \psi \ ,
\end{equation}
while the equation for the CDM perturbations is:
\begin{equation}\label{appB33}
	\a^2 \delta_\cdm'' + \bl( \frac{1 + \frac{3}{2}\a}{1+\a} \br) \a \delta_\cdm' = -\ok^2 \psi + 3\a^2 \psi'' + 3\bl( \frac{1 + \frac{3}{2}\a}{1+\a} \br)\a \psi' \ .
\end{equation}

During the Radiation Domination era the equation for $\psi$ is the same as in the WI limit, and therefore \Eq{appB16} is its solution. Repeating the steps from WI in the $\a \ll 1$ regime (but for $g=0$), we see that $\delta_\cdm$ obeys, in terms of $y \equiv \a / \a_k$:
\begin{equation}\label{appB34}
	y^2 \delta_\cdm'' + y \delta_\cdm' \approx 3y^2 \psi'' + 3 y \psi' -y^2 \psi
\end{equation}
\begin{equation}\label{appB35}
	\Rightarrow \delta_\cdm \xrightarrow{y \gg 1} -\frac{9}{2} \bl( -1 + 2 \gamma_E - \log 3 + 2\log y \br) \ 
\end{equation}
just as in \LC, while $\delta_\idm$ follows the same equation as $\delta_\dr$ in \Eq{appB18}:
\begin{equation}\label{appB36}
	\delta_\idm'' + \frac{1}{3}\delta_\idm \approx 3 \psi'' - \psi \ ,
\end{equation}
\begin{equation}\label{appB37}
	\Rightarrow \delta_\idm \approx \frac{9\bl( y(y^2 - 6)\cos(y/\sqrt{3}) - 2 \sqrt{3}(y^2 - 3)\sin(y/\sqrt{3}) \br)}{2 y^3} \ .
\end{equation}
It can then be seen what was described in the body of the paper: that while $\delta_\cdm$ grows logarithmically, the $\delta_\idm$ tracks the oscillatory behavior of the $\delta_\dr$. This means that only the fraction $1-f$ of DM that is CDM clumps and forms structure, while the remaining $f$ that is IDM does not. This means that after some time $\delta_\idm \ll \delta_\cdm$.

For the slow modes deep inside the Hubble radius we can repeat the steps in the DP with the following changes:
\begin{itemize}
	\item The smallness of $\delta_\idm \ll \delta_\cdm$ (and neglecting baryons) guarantees that the only contribution to the right hand side of \Eq{appB33} comes from the $1-f$ fraction of DM that is CDM:
	\begin{equation}\label{appB38}
		\ok^2 \psi \approx - (1-f) \frac{\frac{3}{2}\a}{1 + \a} \delta_\cdm \ .
	\end{equation}
	As described in subsection \ref{subsec:lims}, this remains the case throughout the rest of the age of the Universe.
	\item It can be shown that $\psi$ is not constant but $\a \psi' \propto f \psi$. Nevertheless, from \Eq{appB38} we know that $\psi \propto \ok^{-2}\delta_\cdm$ and therefore, because $\ok \gg 1$, $\psi'$ is subdominant, and so is $\psi''$.
\end{itemize}
Therefore \Eq{appB33} reduces to a modified version of the M\'{e}sz\'{a}ros equation:
\begin{equation}\label{appB39}
	\a^2 \delta_\cdm'' + \bl( \frac{1 + \frac{3}{2}\a}{1+\a} \br) \a \delta_\cdm' -(1-f)\frac{\frac{3}{2}\a}{1+\a}\delta_\cdm \approx 0 \ ,
\end{equation}
whose solutions are (\cite{Hu:1995en}):
\begin{eqnarray}
	\delta_1 & \equiv & (1+\a)^{-\beta_-} \leftidx{_2}F_1\bl[ \beta_-; \beta_- + \frac{1}{2}; 2 \beta_- + \frac{1}{2}; \frac{1}{1+\a} \br] \label{appB40} \ , \\
	\delta_2 & \equiv & (1+\a)^{-\beta_+} \leftidx{_2}F_1\bl[ \beta_+; \beta_+ + \frac{1}{2}; 2 \beta_+ + \frac{1}{2}; \frac{1}{1+\a} \br] \label{appB41} \ ,
\end{eqnarray}
with $\beta_\pm \equiv \frac{1}{4}\bl( 1 \pm \sqrt{25-24f} \br)$. Note that during MD ($\a \gg 1$), $\delta_1$ is the growing solution; $\delta_1 \propto \a$ for the $f=0$ (\LC) case.

$\delta_{1,2}(\a \ll 1) \sim \mathrm{const.} + \log \a$, and therefore, following Weinberg's method, we find a linear combination that matches \Eq{appB35}.

Taking the ratio of $\delta_\cdm$ in DP to $\delta_\cdm$ in $\LC+\dN$ in the $\a \gg 1$ limit gives

\begin{equation}\label{appB42}
	S(\a)_\mathrm{DP} \xrightarrow{\a \gg 1} \a^{-1-\beta_-}\bl( 0.38 \times \frac{2^{-2\beta_-} \Gamma(2 \beta_-) \tan (2\pi\beta_-)}{\Gamma(2\beta_- + 1/2)} \br) \bl( \frac{-0.05 + 2 \Psi(1-2\beta_-) + \log k \eta_\eq}{-1.9 + \log k \eta_\eq} \br) \ ,
\end{equation}
where $\Psi(x) \equiv \frac{\Gamma'(x)}{\Gamma(x)}$ is the polygamma function of order $0$. For small $f$ we can expand the above result and obtain:
\begin{equation}\label{appB43}
	S(\a \gg 1)_\mathrm{DP} \approx \a^{-\frac{3}{5}f} \bl( 1 + f \bl( \frac{2.0 - 0.57 \log k \eta_\eq}{-1.9 + \log k \eta_\eq} \br) \br) \ .
\end{equation}
Ignoring the (subleading) parenthesis term above and recalling that $a \propto \eta^2$, we find that $\delta_\cdm \sim \eta^{2 - \frac{6}{5}f}$ and
\begin{equation}\label{appB44}
	S(\a \gg 1)_\mathrm{DP} \propto \eta^{-\frac{6}{5}f} \ ,
\end{equation}
as was mentioned in this paper and was derived in \cite{Chacko:2016kgg}.

As stated in the body of the paper, deep in the matter dominated era $\delta_\idm$ begins to grow at a rate equal to that of $\delta_\cdm$, but because by then $\delta_\cdm \gg \delta_\idm$, the IDM contribution to the MPS remains negligible. Therefore, the MPS in DP is suppressed compared to $\LC+\dN$:
\begin{equation}\label{appB45}
	\frac{P(k)_{\mathrm{DP}}}{P(k)_{\LC+\dN}} = \frac{\langle \delta_\ma \delta_\ma \rangle}{\langle \delta_\ma \delta_\ma \rangle\vert_{f=0}} \approx \frac{\langle (1-f)\delta_\cdm (1-f)\delta_\cdm \rangle}{\langle \delta_\cdm \delta_\cdm \rangle\vert_{f=0}} \approx \bl[ (1-f) S(\a \gg 1)_\mathrm{DP} \br]^2
\end{equation}
which again can be rewritten as (\cite{Chacko:2016kgg}):
\begin{equation}\label{appB46}
	\frac{P(k)_{\mathrm{DP}}}{P(k)_{\LC+\dN}} \approx (1-2f) \bl( \frac{\eta}{\eta_\eq} \br)^{-\frac{12}{5}f}
\end{equation}
to leading order in $f$, having used \Eq{appB44}.


\bibliography{ids}{}

\end{document}